\documentclass[11pt,a4paper]{article}
\pdfoutput=1
\usepackage{amsmath,amssymb,dsfont}
\usepackage{graphicx,color}
\usepackage{verbatim, cancel, bbm}
\usepackage[bf]{caption}
\usepackage{ulem,stackrel}
\usepackage{epsfig}

\usepackage{epsfig}

\usepackage[breaklinks,bookmarks=true,colorlinks=true,linkcolor=blue,citecolor=blue,urlcolor=blue,hyperfigures=true]{hyperref}
\usepackage[nosort]{cite} 
\usepackage[bulletsep]{collref} 

\numberwithin{equation}{section}

\ifx\href\asklfhas\newcommand{\href}[2]{#2}\fi

\definecolor{pink}{rgb}{0.7,0,0.7}
\definecolor{green}{rgb}{0,0.5,0}
\definecolor{orange}{rgb}{1,0.4,0.3}

\newcommand{\be}{\begin{equation}}
\newcommand{\ee}{\end{equation}}
\newcommand{\ba}{\begin{aligned}}
\newcommand{\ea}{\end{aligned}}
\newcommand{\ben}{\begin{displaymath}}
\newcommand{\een}{\end{displaymath}}
\newcommand{\bea}{\begin{eqnarray}}
\newcommand{\eea}{\end{eqnarray}}
\newcommand{\bean}{\begin{eqnarray*}}
\newcommand{\eean}{\end{eqnarray*}}
\newcommand{\bpmat}{\begin{pmatrix}}
\newcommand{\epmat}{\end{pmatrix}}

\newcommand{\AdS}{{\ensuremath{\mathrm{AdS}}}}
\newcommand{\Sph}{{\ensuremath{\mathrm{S}}}}

\newcommand{\PSU}{{\ensuremath{\mathrm{PSU}}}}
\newcommand{\SO}{{\ensuremath{\mathrm{SO}}}}
\newcommand{\SLS}{{\ensuremath{\mathrm{SL}}}}
\newcommand{\SU}{{\ensuremath{\mathrm{SU}}}}

\newcommand{\OSP}{{\ensuremath{\mathrm{OSP}}}}

\newcommand{\alg}[1]{\ensuremath{\mathfrak{#1}}}

\newcommand{\psu}{{\ensuremath{\mathfrak{psu}}}}
\newcommand{\so}{{\ensuremath{\mathfrak{so}}}}

\newcommand{\osp}{{\ensuremath{\mathfrak{osp}}}}

\newcommand{\Reals}{{\ensuremath{\mathbb{R}}}}

\newcommand{\calN}{{\ensuremath{\mathcal{N}}}}

\newcommand{\calC}{{\ensuremath{\mathcal{C}}}}

\renewcommand{\l}{\lambda}

\renewcommand{\a}{\alpha}
\renewcommand{\b}{\beta}
\newcommand{\g}{\gamma}
\newcommand{\G}{\Gamma}
\renewcommand{\d}{\delta}
\newcommand{\eps}{\epsilon}

\newcommand{\s}{\sigma}

\newcommand{\m}{\mu}
\newcommand{\n}{\nu}

\renewcommand{\o}{\omega}

\newcommand{\f}{\frac}

\newcommand{\la}{\langle}

\newcommand{\ra}{\rangle}

\renewcommand{\tt}{\ensuremath{\mathbf{t}}}

\newcommand{\dd}{{\ensuremath{\text{d}}}}

\long\def\symbolfootnote[#1]#2{\begingroup
\def\thefootnote{\fnsymbol{footnote}}\footnote[#1]{#2}\endgroup}

\begin{document}

\begin{titlepage}
\hfill\parbox[t]{4cm}{
\texttt{FU-Ph 10/2016~(07) \\
NORDITA-2016-110 \\
ZMP-HH/16-24}
}

\vspace{20mm}

\begin{center}

{\Large\bf  Coset construction of $\rm AdS$ particle dynamics}

\vspace{30pt}

{Martin Heinze,$^{a,\,b}~$ George Jorjadze,$^{c,\,d}~$ Luka Megrelidze,$^e$
}
\\[6mm]

{\small
{\it\ ${}^a$II. Institut f{\"u}r Theoretische Physik, Universit{\"a}t Hamburg,\\
	Luruper Chaussee 149, 22671 Hamburg, Germany}\\[2mm]
{\it\ ${}^b$Zentrum f\"ur Mathematische Physik, Universit\"at Hamburg,\\ 
	Bundesstrasse 55, 20146 Hamburg, Germany}\\[2mm]
{\it${}^c$Free University of Tbilisi,\\
		Agmashenebeli Alley 240, 0159, Tbilisi, Georgia}\\[2mm]
{\it${}^d$Razmadze Mathematical Institute of TSU,\\
Tamarashvili 6, 0177, Tbilisi, Georgia}\\[2mm]
{\it${}^e$Ilia State University,\\
K. Cholokashvili Ave 3/5, 0162, Tbilisi, Georgia}\\[5mm]
\texttt{martin.heinze@desy.de},\qquad\texttt{george.jorjadze@physik.hu-berlin.de}\\
\texttt{luka.megrelidze.1@iliauni.edu.ge},
}

\vspace{45pt}

\end{center}

\centerline{{\bf{Abstract}}}
\vspace*{5mm}
\noindent
	We analyze dynamics of the AdS$_{N+1}$ particle realized on the coset SO$(2,N)/$SO$(1,N)$. 
	Hamiltonian reduction provides the physical phase space in terms of the coadjoint orbit obtained by boosting a timelike element of $\frak{so}(2,N)$. We show equivalence of this approach to geometric quantization and to the $\SO(N)$ covariant oscillator description, for which the boost generators entail a complicated operator ordering.
	As an alternative scheme, we introduce dual oscillator variables and derive their algebra at the classical and the quantum level.
	This simplifies the calculations of the commutators for the boost generators and leads to unitary irreducible representations of $\frak{so}(2,N)$ for all admissible values of the mass parameter.
	We furthermore discuss an $\SO(N)$ covariant supersymmetric extension of the oscillator quantization, with its realization for superparticles in AdS$_2$ and AdS$_3$ given by recent works.

\vspace{4cm}

\end{titlepage}


\section{Introduction and Conclusion}\label{sec:Intro}
	During the last decade the $\AdS$/CFT correspondence and its connection to integrable models have immensely improved our understanding of both conformal field theories (CFT) as well as string theories in Anti-de Sitter space ($\AdS$), see the reviews \cite{Arutyunov:2009ga, Beisert:2010jr, Bombardelli:2016rwb}. To test this conjectured duality one major task is to compare the CFT conformal dimensions with the string energy spectrum $E$. In particular, the best studied duality pair, the correspondence between $\calN=4$ super Yang-Mills theory and the $\AdS_5\times\Sph^5$ superstring, appears to be quantum integrable in the planar limit, which allowed for a solution of the spectral problem through the mirror Thermodynamic Bethe Ansatz (TBA) \cite{Arutyunov:2009ur, Bombardelli:2009ns, Gromov:2009tv, Gromov:2009bc} as well as the Quantum Spectral Curve \cite{Gromov:2013pga}.

	One might hope that this impressive progress would elucidate how quantization of $\AdS$ string theories can be achieved. However, even a derivation of the $\AdS_5 \times \Sph^5$ superstring spectrum from first principles is still an open problem. The spectrum of the corresponding supergravity, {\it viz.} the $1/2$-BPS sector of the superstring, is well-known and has been shown to match with the spectrum of the {\it massless} $\AdS_5 \times \Sph^5$ superparticle \cite{Horigane:2009qb}, see also \cite{Metsaev:1999gz, Metsaev:2000yu, Siegel:2010gm} as well as the recent work on the supertwistor formulation \cite{Arvanitakis:2016vnp}.

	To obtain the spectrum of more-involved string states, since the seminal works \cite{Drukker:2000ep, Frolov:2002av, Berenstein:2002jq, Gubser:2002tv, Frolov:2003qc, Arutyunov:2003uj} it has been a standard approach to quantize semi-classical string solutions as well as the fluctuations around them.
	This technique is however only valid if some of the $\psu(2,2|4)$ charges diverge in 't Hooft coupling as $\sqrt{\l}$, rendering the string state {\it heavy}, $E \propto \sqrt{\l}$. For {\it light} string states expansion of the Lagrangian formally breaks down and it is a renowned challenge to obtain the spectrum beyond the leading order, $E \propto \l^{1/4}$ \cite{Gubser:2002tv}.
	These difficulties can be traced back to the particular behavior of the center-of-mass degrees of freedom \cite{Passerini:2010xc} as well as incompatibility of customary light-cone gauge choices.

	In \cite{Frolov:2013lva} an alternative scheme has been applied to a pulsating string solutions in bosonic $\AdS_5\times\Sph^5$. Using static gauge \cite{Jorjadze:2012iy}, see also \cite{Heinze:2015xxa}, 
	it was shown that the calculation of the string energy spectrum reduces to the problem of {\it massive} $\AdS_5$ particle dynamics, with the mass spectrum determined by internal string degrees of freedom.
	Allowing heuristically for supersymmetric effects then provided an operator ordering, resulting in a match of the string energy with the Konishi anomalous dimension to the first quantum corrections of order $\l^{-1/4}$.
	
	As \cite{Frolov:2013lva} describes the $\SO(2,4)\times\SO(6)$ isometry group orbit of a pulsating string it appears promising to apply the gauge invariant Kirillov-Kostant-Souriau method of coadjoint orbits also to other string solutions. 
	In \cite{Heinze:2014cga} we followed this proposal and quantized the bosonic $\AdS_3\times\Sph^3$ spinning string solutions of \cite{Frolov:2003qc, Arutyunov:2003uj}, yielding known results for the special case of bosonic point particles \cite{Dzhordzhadze:1994np} as well as correct asymptotics for extended spinning strings. 
	To acquire a match with the integrability based predictions beyond the leading order it became clear that supersymmetry has to be treated properly.
	
	Therefore, as a first example for quantization of superisometry group orbits, the coadjoint orbit method was applied to the $\AdS_2$ superparticle on $\OSP(1|2)/\SO(1,1)$ \cite{Heinze:2015oha}. The resulting Noether charges formed a Holstein-Primakoff-like realization of $\osp(1|2)$, where for the massless case however $\kappa$-symmetry left an insufficient amount of fermions to quantize the model.

	Recently, this problem has been circumvented by studying the $\AdS_3$ superparticle on $\OSP(1|2)\times\OSP(1|2)/\SLS(2,\Reals)$ \cite{Heinze:2016fin}. Here, calculation of the symplectic form as well as of the Noether charges naturally split up into left and right chiral sectors, yielding a quantum realization of $\osp_l(1|2)\oplus\osp_r(1|2)$.
	For the massless particle it was then found that the superisometry algebra extends to the corresponding superconformal algebra $\osp(2|4)$, with its 19 charges realized by all possible real quadratic combinations of the phase space variables.

	We motivated the above program with the goal to understand quantization of the type IIB superstring on $\AdS_5 \times \Sph^5$ and semi-classical string states thereof. This is however not the only area of application. Currently, the machinery developed for $\AdS_5 \times \Sph^5$ is adapted to integrable string theories in less supersymmetric spaces \cite{Zarembo:2010sg}, that is besides from $\AdS_4 \times \mathrm{CP}^3$ \cite{Klose:2010ki} also to $\AdS_2 \times \Sph^2 \times \mathrm{M}^6$ \cite{Sorokin:2011rr, Cagnazzo:2011at} 
	and $\AdS_3 \times \Sph^3 \times \mathrm{M}^4$ \cite{David:2008yk, Babichenko:2009dk}, see also the review \cite{Sfondrini:2014via}. 
	Therefore, the works \cite{Heinze:2015oha} and \cite{Heinze:2016fin} are naturally viewed as truncations of the superstrings in the latter two backgrounds and it seems appealing to generalize them to $\AdS_2$ and $\AdS_3$ superparticles on the corresponding cosets build from the supergroup $\SU(1,1|2)$, $\PSU(1,1|2)$, and more generally $\mathrm{D}(2,1;\a)$, see also the related works \cite{Galajinsky:2010zy, Galajinsky:2011xp, Bellucci:2011hk} and \cite{Krivonos:2010zy, Kozyrev:2013vla, Kozyrev:2016mlo, Galajinsky:2016wuc} and references therein. 
	
	In the present work we are though following an orthogonal direction of research. Instead of increasing the amount of supersymmetry another possible generalization is to raise the dimension of the $\AdS$ space.
	Our main focus will be on the massive bosonic $\AdS_{N+1}$ particle. Although quantization of this system has been achieved before \cite{Fronsdal:1974ew, Breitenlohner:1982jf, Aharony:1999ti, Dorn:2005jt, Dorn:2005ja, Dorn:2010wt, Jorjadze:2012jk},
	the realization of the supersymmetric backgrounds \cite{Zarembo:2010sg} in terms of semi-symmetric spaces demands that quantization should be understood in a scheme corresponding to the coset description $\AdS_{N+1}=\SO(2,N)/\SO(1,N)$.

	We will start by applying the Kirillov-Kostant-Souriau method to particle dynamics on general bosonic cosets $G/H$, where we will show explicitly how gauge invariant Hamiltonian reduction gives rise to the physical phase space.
	For the restriction to $\AdS_{N+1}=\SO(2,N)/\SO(1,N)$ the physical phase space of the massive particle is given by the coadjoint orbit obtained by boosting a temporal element of $\so(2,N)$. By this, we rederive in a succinct manner the results of geometric quantization \cite{Dorn:2005jt} and the oscillator description \cite{Dorn:2005ja}, where the $\SO(2,N)$ charges $J_{AB}$ take a form manifestly covariant under the spatial subgroup $\SO(N) \subset \SO(2,N)$ \cite{Heinze:footnote1}. 
	
	Additionally, we propose an alternative quantization scheme in terms of a set of dual oscillator variables, which in comparison to \cite{Dorn:2005ja} simplifies the calculation of the commutators of boost generators. This simplification can be attributed to the algebra of these dual oscillators, which we derive at the classical as well as at the quantum level. To our knowledge, this algebra has not been discussed previously in the literature.

	With the goal to understand orbit method quantization of particles on the supersymmetric backgrounds \cite{Zarembo:2010sg} it seems promising to try to find a supersymmetric extension of the found $\so(2,N)$ algebra which preserves the spatial $\SO(N)$ covariance.
	At the end of this work, we therefore revisit the results for the $\calN=1$ $\AdS_2$ \cite{Heinze:2015oha} and $\AdS_3$ \cite{Heinze:2016fin} superparticles and rewrite them in the desired $\SO(N)$ covariant form. This yields an ansatz for the supercharges of an $\calN=1$ $\AdS_{N+1}$ superparticle. We observe however that due to a certain constraint on the $\SO(2,N)$ gamma matrices, or equivalently due to absence of an $R$-symmetry, this ansatz can only be consistent for $N\leq3$. 
	While the cases $N=1$ and $N=2$ are covered by \cite{Heinze:2015oha} and \cite{Heinze:2016fin}, showing consistency for $N=3$ is still an open problem, even at the classical level. It would be interesting to compare the obtained ansatz for the $\calN=1$ $\AdS_4$ superparticle with the supertwistor formulation \cite{Arvanitakis:2016vnp}.

	An idea similar to the above extension has been adopted in \cite{Galajinsky:2016wuc}, where the dynamical realization on $\SU(1,1|2)$ has been generalized to $\SU(1,1|N)$. Furthermore, if \cite{Heinze:2016fin} can be extended to $\AdS_3$ superparticles with a higher amount of supersymmetry, generalization of its spatial $\SO(2)$ to an $\SO(N)$ covariance should again lead to an ansatz for higher dimensional $\AdS$ superparticles with the same amount of supersymmetry. 
	As the $\calN=1$ supersymmetric extension in this work limits to the bosonic charges as discussed in \cite{Dorn:2005jt, Dorn:2005ja}, for other $\AdS_{N+1}$ superparticles the alternative quantization scheme in terms of dual oscillators might prove useful.

	Such $\AdS$ superparticles will have a non-vanishing $R$-symmetry. Hence, another congenial generalization of the present work would be to study orbit method quantization for the bosonic $\AdS_{N+1} \times \Sph^M$ particle realized on the coset $(\SO(2,N)/\SO(1,N))\times(\SO(M+1)/\SO(M))$.
	Preliminary studies show that the physical phase space of this system is realized not by a single orbit but rather by a family of orbits parametrized by a canonical pair 'shifting' mass/momentum form $\AdS$ to the sphere, which resembles the setting of the BMN string \cite{Berenstein:2002jq}.

	The paper is organized as follows. In Section \ref{sec:GHcoset} we apply the method of coadjoint orbits to general bosonic cosets $G/H$, which is then restricted to the case of $\SO(2,N)/\SO(1,N)$ in Section \ref{sec:SOcoset}. In Section \ref{sec:DualOsc} we propose an alternative quantization scheme in terms of dual oscillators.
	Finally, in Section \ref{sec:Super} we discuss a supersymmetric extension of the coset scheme and its realization for $\AdS_2$ and $\AdS_3$. Some technical details are collected in four appendices.

\section{\texorpdfstring{$G/H$}{G/H} Coset Model}\label{sec:GHcoset}

	Let $G$ be a semi-simple Lie group and $H$ its semi-simple subgroup with dimensions $N_G$ and $N_H$, respectively.
	We denote by $\alg{g}$ and $\alg{h}$ the corresponding Lie algebras and we also introduce $\alg{h}_\perp$ as the orthogonal completion of $\alg{h}$ in $\alg{g}$.

	The commutation relations of the basis vectors of $\alg{g}$
	\be\label{Lie algebra}
		[\tt_a, \tt_b]=f_{ab}{}^{c}\,\tt_c
	\ee
	define the structure constants which provide a non-degenerated Killing form with metric tensor
	$\rho_{ab}:=\la \tt_a \,\tt_b \ra= f_{ad}{}^{c}f_{bc}{}^{d}$.
	One can choose a basis where $\tt_\a$, with $\a=(1, \dots, N_H$), form a basis of $\alg{h}$ and the remaining $N_G - N_H$ elements $\tt_{\bar\a}$,
	with $\bar\a=(N_H +1, \dots ,  N_G)$, belong to $\alg{h}_\perp$. Expanding $V\in \alg{g}$ in this basis, $V=V^\a\,\tt_\a+ V^{\bar\a}\,\tt_{\bar\a}$, one obtains
	\be\label{<s s>}
		V^{\bar\a}  V_{\bar\a}=\la V\,V \ra- V^\a  V_{\a}~,
	\ee
	with $V_\a=\rho_{\a\b}V^\b$ and $V_{\bar\a}=\rho_{\bar\a\bar\b}V^{\bar\b}$. 
	Due to \eqref{<s s>}, $V^{\bar\a}  V_{\bar\a}$
	is invariant under the adjoint transformations
	$V\mapsto h \,V\, h^{-1}$, with $h\in H$.

	The dynamics of a particle in the coset construction is defined by the action
	\be\label{coset action 0}
	S=\int \dd \tau \left(\f{\la (\dot g\,g^{-1}-{\cal A})^2\rangle}{2e} -\frac{em^2}{2}\right)~,
	\ee
	where $g\in G$, ${\cal A}\in \alg{h}$, $m$ is the particle mass and $e$ plays the role of einbein.
	This action is invariant under the worldline reparameterization $ \tau \mapsto\tilde\tau=\varphi(\tau),$ 
	together with
	\be\label{reparametrization}
		g(\tau)\mapsto \tilde g(\varphi)=g(\tau)~, \quad {\cal A}(\tau) \mapsto \tilde{\cal A}(\varphi)=
	{\cal A}(\tau)/\dot\varphi(\tau)~, \quad e(\tau)\mapsto \tilde e(\varphi)=e(\tau)/\dot\varphi(\tau)~.
	\ee
	Another gauge symmetry of \eqref{coset action 0} is given by the local transformation
	\be\label{local left symmetry}
		g(\tau) \mapsto h(\tau)\, g(\tau)~, \qquad {\cal A}(\tau) \mapsto h(\tau){\cal A}(\tau) \,h^{-1}(\tau)
		+\dot h(\tau)\,h^{-1}(\tau)~,
	\ee
	with $h(\tau)\in H$. The right multiplication $ g\mapsto g\, f$, with $f\in G$ being  $\tau$-independent, is a global symmetry of the action \eqref{coset action 0} and the corresponding Noether charge reads
	\be\label{Noether charge R}
		R=\f{g^{-1}\dot g\,-g^{-1}{\cal A}\, g}{e}~.
	\ee

	Expanding $\cal A$ in the basis ${\cal A}={\cal A}^\a\tt_\a$ and varying \eqref{coset action 0} with respect to ${\cal A}^\a$ we find
	${\cal A}^\a=V^\a$, where $V= \dot g\,g^{-1}$ is the Maurer-Cartan form.
	The insertion of this ${\cal A}^\a$ back in \eqref{coset action 0}
	leads to the gauge invariant action written solely in terms of $g$, such that by \eqref{<s s>} 
	we end up with
	\be\label{coset action 1}
	S=\int \dd \tau \left(\f{V^{\bar\a}V_{\bar\a}}{2e} -\frac{e m^2}{2}\right)~.
	\ee
	The invariance of this action under the gauge transformations $g(\tau) \mapsto h(\tau)\, g(\tau)$ follows from 
	the orthogonality conditions
	$\la \tt_{\a}\,\tt_{\bar\a}\ra=0$ and
	from the above mentioned invariance of the right hand side of \eqref{<s s>} under the adjoint action of $H$.

	In the first order formalism the action \eqref{coset action 0} is equivalent to
	\be\label{coset action 2}
	S=\int \dd \tau \Big[\langle\, L\,\dot g\,g^{-1}\,\rangle-\frac{e}{2}\left(\langle\, L\,L\,\rangle+m^2\right)
	-\langle\, {\cal A} L\,\rangle \Big]~,
	\ee
	with $L\in \alg{g}$. Note that the variation of $L$ in \eqref{coset action 2} yields the equation
	\be\label{L=dot g}
	eL=\dot g\,g^{-1}-{\cal A}~,
	\ee
	and its insertion back in \eqref{coset action 2} reproduces the initial action \eqref{coset action 0}.
	This shows the equivalence between the actions \eqref{coset action 0} and \eqref{coset action 2}.
	Equation \eqref{L=dot g} also relates $L$ and $R$ by
	\be\label{L-R orbit}
	R=g^{-1}\,L\,g~.
	\ee

	Writing the last term of \eqref{coset action 2} as ${\cal A}^\a L_\a$,
	with $L_\a=\langle\, \tt_\a \,L\,\rangle,$
	we find that ${\cal A}^\a$ play the role of Lagrange multipliers, as does the einbein $e$. Their variations provide the constraints
	\be\label{Constaints 2}
		\la L \,L \ra+m^2=0~, \qquad L_\a=0~.
	\ee

	Thus, one gets a constraint dynamical system on $T^*G$ with phase space variables ($g, \,L$).
	The symplectic form $\dd \langle\, L\,\dd  g\,g^{-1}\,\rangle$ leads to the Poisson brackets
	\be\ba\label{Basic PB in T*G}
		&\{g, g\}=0~,& \quad &\{L_a, g\}=t_a\,g~,&  \quad &\{L_a, L_b\}=-f_{ab}{}^c\,L_c~,&  \\[1mm]
		&\{R_a, g\}=g\,t_a~,&   &\{L_a, R_b\}=0~,&  &\{R_a, R_b\}=f_{ab}{}^c\,R_c~,&
	\ea\ee
	with $L_a=\langle\, t_a \,L\,\rangle$ and $R_a=\langle\, t_a \,L\,\rangle$.
	Note that the structure constants with down indices $f_{abc}= f_{ab}{}^{d}\,\rho_{dc}$ are completely antisymmetric,
	which yields $\{\la L \,L \ra, L_a\}=0.$
	Hence, \eqref{Constaints 2} defines the first class constraints and they generate the gauge transformations
	\eqref{reparametrization}-\eqref{local left symmetry}.

	The dynamical integrals $R_a$ are gauge invariant since they have vanishing Poisson brackets with the constraints.
	Another set of gauge invariant variables are given by the coordinates on $G/H$.
	A complete set of gauge invariant variables defines the physical phase space and
	the Hamiltonian reduction to this space prepares the system for quantization. 
	After quantization the operators $R_a$ have to form a unitary representation of $G$. 
	
	Another possibility to simplify the reduction procedure is to use gauge fixing.
	For this, one has to impose $N_H+1$ conditions to fix the gauge completely, which
	defines a $2(N_G - N_H-1)$ dimensional physical phase space. 
	Details of the reduction procedure essentially depends on the structure of $G$ and $H$.

	In some cases the gauge freedom allows to choose $L$ in the form $L=m\tt_0$,
	where $\tt_0$ is a fixed unit element of $\alg{h}_\perp$.
	The dynamical integrals $R_a$ are then on the coadjoint orbit of $m\tt_0$
	and by \eqref{coset action 2} and \eqref{L-R orbit} one gets the Kirillov-Kostant symplectic form
	\be\label{KK-form}
	\o=m\,\dd \la \tt_0\,\dd g\,g^{-1}\ra~,
	\ee
	which leads to unitary irreducible representations of $G$.

	In the next section we consider the case $G=\SO(2,N)$ and $H=\SO(1,N)$, for which \eqref{KK-form} holds.
	The corresponding coset model $G/H$ describes the dynamics of a particle in $\AdS_{N+1}$ and leads to the high weight
	unitary irreducible representation of $\SO(2,N)$.

\section{\texorpdfstring{$\SO(2,N)/\SO(1,N)$}{SO2n/SO1N} Coset Model}\label{sec:SOcoset}

	Let us consider $\Reals^{2,N}$ with coordinates $X^{A}$, $A=0', 0, 1,\dots ,N$, and the metric tensor $\eta_{{AB}}=\mbox{diag}(-1, -1, 1,\dots ,1)$.
	The linear isometry transformations of $\Reals^{2,N}$ are given by $X^A\mapsto g^A_{\,~B}\,X^B$, where the matrix $g^A_{\,~B}\in \SO(2,N)$ fulfills the identity
	\be\label{SO(2,N) conditions}
		g^{A}_{\,~C}\,g_{B}^{\,~C}=\d^A_B=g_C^{\,~A}\,g^C_{\,~B}~,
	\ee
	with $g_{A}^{\,~B} = \eta_{AC}\,\eta^{BD} g^{C}_{\,~D}$. Using the infinitesimal form $g^A_{\,~B}=\d^A_{B}+\eps^A_{\,~B}$, one finds
	$\eps^{AB}+\eps^{BA}=0$ and by this skew-symmetric property one gets
	$\eps^A_{\,~B}=\frac{1}{2}\,\eps^{CD}(\tt_{CD})^A_{~B}\,$, with
	\be\label{so(2,N) basis}
	(\tt_{CD})^A_{~B}=\d_C^A\, \eta_{BD}-\d_D^A\, \eta_{BC}~.
	\ee
	Here, $A$ and $B$ are matrix indices while ${CD}$ enumerate matrices
	forming a basis of $\so(2,N)$, {\it i.e.} they satisfy the commutation relations
	\be\label{so(2,N) matrix algebra}
	[\tt_{AB}, \tt_{CD}]=\eta_{AD}\tt_{BC}+\eta_{BC}\tt_{AD}
	-\eta_{AC} \tt_{BD}-\eta_{BD}\tt_{AC}~.
	\ee
	The matrices $\tt_{0'k}$ and $\tt_{0k}$, for $k=1,\dots , N$, generate boosts, while the generators of the compact subgroup $\SO(2)\times\SO(N)$ are $\tt_{0'0}$ and $\tt_{kl}$.

	The inner product in $\so(2,N)$ introduced by the normalized Killing form,
	\be\label{Norm-Trace}
		\la \tt_{AB} \, \tt_{CD}\ra=
		\frac{1}{2}\,\mathrm{Tr}\left(\tt_{AB}\, \tt_{CD}\right)=\eta_{AD}\eta_{BC}-\eta_{AC}\eta_{BD}~, 
	\ee
	provides the following norms of the basis vectors 
	\be\label{norm 1}
		\la \tt_{0'k} \, \tt_{0'k}\ra=1=\la \tt_{0k} \, \tt_{0k}\ra ~, \qquad
		\la \tt_{0'0} \, \tt_{0'0}\ra=-1=\la \tt_{kl} \, \tt_{kl}\ra~.
	\ee
	Here, $k\neq l$ and we assume no summation over the indices $k$ and $l$. Note that two different basis vectors are orthogonal to each other, {\it i.e.} $\tt_{AB}$ form an orthonormal basis.
	Then the inner product of two elements $U=\frac{1}{2}\, U^{AB} \tt_{AB}$ and $V=\frac{1}{2}\, V^{AB} \tt_{AB}$  is given by $\langle U\,V\rangle=\frac{1}{2}\,U^{AB} V_{BA}$ and $\la \tt_{AB} \, V \ra=V_{BA},$
	where $V_{BA}=\eta_{BC}\eta_{AD}V^{CD}$.

	We choose the $\SO(1,N)$ subgroup by the matrices with the block structure
	\be\label{SO(1,N) subgroup}
		h=\begin{pmatrix}
		1 & \\
		 & \Lambda^\m_{~\,\n}
		\end{pmatrix},
	\ee
	where we left the vanishing upper-right $1\times N$ and lower-left $N \times 1$ blocks blank and $\Lambda^\m_{~\n}$ are Lorentz matrices with $\m,\n =0,1,\dots, N$.
	The matrices $\tt_{\m\n}$ then form a basis of the $\so(1,N)$ subalgebra
	and the gauge potential can be written as ${\cal A}=\frac{1}{2}\, {\cal A}^{\m\n}\tt_{\m\n}$.

	The gauge invariant action \eqref{coset action 0} then becomes
	\be\label{AdS coset action 0}
		S=\int \dd \tau \left(\f{\la\,(\dot g\,g^{-1})^2\,\ra-{\cal A}^{\m\n}
		\la\,\tt_{\m\n}\,\dot g\,g^{-1}\ra+\f{1}{2}\,{\cal A}^{\m\n}{\cal A}_{\n\m}}{2e} -\frac{em^2}{2}\right)~,
	\ee
	and after the elimination of ${\cal A}^{\m\n}$ by ${\cal A}_{\m\n}=-\la \tt_{\m\n}\,\dot g\,g^{-1}\ra$, 
	we find the action  \eqref{coset action 1}
	as
	\be\label{AdS coset action 1}
		S=\int \dd \tau \left(\f{\la  \tt_{0'}^{~\,\,\m} \, \dot g\,g^{-1} \ra\,\la  
		\tt_{0'\m} \, \dot g\,g^{-1}\ra}{2e}-\frac{em^2}{2}\right)~.
	\ee
	Using then \eqref{SO(2,N) conditions}-\eqref{so(2,N) basis}, one gets
	$\la \tt_{0'\m}\, \dot g\,g^{-1}\ra=\dot g^{0'}_{~\,B}\,g_\m^{~\,B}$ and \eqref{AdS coset action 1} reduces to
	\be\label{AdS coset action 1-2}
		S=\int \dd \tau \left(\f{\dot x^A\,\dot x_A}{2e}-\frac{em^2}{2}\right)~,
	\ee
	with $x_A=g^{0'}_{\,~A}$. These components form the first row of the matrix $g^A_{~\,B}$ and they are obviously invariant
	under the gauge transformations $g\mapsto h\,g$.
	The coordinates $x^A$ are constrained to lie on the hyperboloid
	$x^Ax_A+1=0$. Hence, the action \eqref{AdS coset action 1-2} describes the dynamics of a particle in
	$\AdS_{N+1}$, see also Appendix \ref{app:AdSNpl1}.

	In the first order formalism one gets $L=L^\m\,\tt_{0'\m}$, with $L^\m L_\m+m^2=0$.
	Its gauge transformations  $L\mapsto h\,L\,h^{-1}$
	transform the components $L^\m=\la\tt_{0'}^{~\,\m}\, L\ra$  by the Lorentz matrices $L^\m\mapsto \Lambda^\m_{~\n}L^\n$
	and they can be turned to $L^\m=m\, \d^\m_0$. Therefore, we can set
	\be\label{fixed L}
		L=m \,\tt_{0'0}~.
	\ee

	Taking for the group element $g=g_c\,g_b$, where $g_c$ is a compact element and $g_b$ is a boost,
	\be\label{decompositiom}
		g_c=e^{\varphi\, \tt_{0'0}}\,e^{\frac{1}{2}\,\phi_{kl} \tt_{kl}}~,  \qquad\qquad
		g_b=e^{\zeta^{0'k} \,\tt_{0'k}+\zeta^{0k}\, \tt_{0k}}~,
	\ee
	we find that the symplectic form \eqref{KK-form} and the Noether charge \eqref{L-R orbit} depend only on $g_b$
	\be\label{reduction 0}
		\o=m\dd \la\tt_{0'0}\, \dd g_b\,g_b^{-1}\ra~, \qquad\qquad   R=m\, g_b^{-1}\,\tt_{0'0}\, g_b~.
	\ee
	Introducing then the components $J_{AB}=\la\tt_{BA}\,R\ra$ and using  \eqref{so(2,N) basis} for $t_{0'0}$,
	we obtain
	\be\label{reduction 1}
		\quad\o=m\,\dd u_{C}\wedge \dd v^{C}~, \qquad\qquad J_{AB}=m\left(u_{A}v_{B}-u_{B}v_{A} \right)~,
	\ee
	where $u_A= (g_b)_{0A}$ and $v_A= (g_b)^{0'}_{~\,\,A}$. 

	These $2(N+2)$ variables are still constrained by the conditions
	\be\label{constraints 3}
	v_A\,v^A+1=0~,  \qquad u_A\,u^A+1=0~, \qquad u_A\, v^A=0~, \qquad u_{0'}+v_{0}=0~,
	\ee
	since $g_b$ is a boost (see Appendix \ref{app:CosetParam}). 
	These four constraints then define a $2N$ dimensional physical phase space, where one can choose $u_k$ and $v_k$, with $k=1,\dots , N$, as global coordinates. 
	Counting the degrees of freedom as described in the previous section, we observe that this matches the dimension of the physical phase space, which is also $2N$.

	To get a convenient parametrization of the Noether charges, we introduce the matrix formed
	by the upper-left $2\times2$ block of $(g_b)^A_{\,~B}$,
	\be\label{C}
		C=\begin{pmatrix}
		\a & \b\\
		\b & \g
		\end{pmatrix}~,
	\ee
	with $\a=(g_b)^{0'}_{\,\,~0'}$, $\b=(g_b)^{0'}_{\,\,~0}=(g_b)^{0}_{\,\,~0'}\,$, $\g=(g_b)^{0}_{\,\,~0}\,$.
	Equation \eqref{constraints 3} is then equivalent to
	\be\label{constraints 4}
		\a^2+\b^2=1+v^2~, \qquad  \g^2+\b^2=1+u^2~, \qquad \a\,\b+\b\,\g=-u\,v~,
	\ee
	with $u^2=u_n u_n$, $v^2=v_n v_n$, $u\,v=u_n v_n$. In Appendix \ref{app:CosetParam} we show that the matrix $C$ is
	semi-positive and, therefore, that the system \eqref{constraints 4} has an unique solution for $\a$, $\b$ and $\g$.

	From \eqref{reduction 1} we obtain a parametrization of the Noether charges in terms of the coordinates $u_k$ and $v_k$,
	\begin{align}\label{Noether charges}
		\qquad&J_{0'0}=m\,\d~,&    &J_{kl}=m(u_k v_l-u_l v_k)~,\qquad\\ \label{boosts 3}
		&J_{k0'}=m(\a\,u_k+\b\,v_k)~,&  &J_{k0}=m(\b\,u_k+\g\,v_k)~,
	\end{align}
	where $\d=\a\,\g-\b^2$ is the determinant of the matrix \eqref{C}.

	Introducing now the $\mathrm{O}(N)$ scalar combinations of the Noether charges
	\be
		\label{scalarDefs}
		J^2 \equiv \f 1 2\, J_{ln} J_{ln}~,\quad
			J_{0'}^2 \equiv J_{n0'} J_{n0'}~,\quad
			J_{0}^2 \equiv J_{n0} J_{n0}~,\quad
			J_{0'}\cdot J_{0}\equiv J_{n0'} J_{n0}~,
	\ee
	and using \eqref{constraints 4}, we find 
	\begin{align}
		\label{scalars}
		&\qquad\qquad\qquad\qquad\qquad\quad J^2=m^2(\d^2-\a^2-2\b^2-\g^2+1)~,\\
		\label{scalars1}
		&J^2_{0'}=m^2(\d^2-\a^2-\b^2)~, \quad J^2_{0}=m^2(\d^2-\g^2-\b^2)~, \quad J_{0'}\cdot J_{0}=-m^2(\a+\g)\b~.
	\end{align}

	From these equations follows that the quadratic Casimir of $\SO(2,N)$, $C_2 \equiv \f 1 2\, J_{AB}\,J^{AB}$, 
	satisfies  the mass-shell condition
	\be\label{mass-shell}
	C_2=
	E^2+J^2-J^2_{0'}-J^2_{0}=m^2\,,
	\ee
	where $E=J_{0'0}$ is the particle energy.

	According to the general scheme described in the previous section, the Poisson brackets of the Noether charges satisfy the $\so(2,N)$ algebra
	\begin{equation}\label{PB_J}
		\{J_{AB},\,J_{CD}\}= \eta_{AC}J_{BD}+\eta_{BD}J_{AC}-
		\eta_{AD}J_{BC}-\eta_{BC}J_{AD}~,
	\end{equation}
	and $J_{kl}$ provide its $\so(N)$ subalgebra. Introducing the complex combinations of the boosts,
	\be\label{B,B*}
		B_k=J_{k0'}-i\,J_{k0}~,  \qquad\qquad B^*_k=J_{k0'}+i\,J_{k0}~,
	\ee
	the remaining Poisson brackets take the form
	\be\ba\label{PB_Bn}
		&\{E,\,J_{kl}\}=0~,& ~~~~~&\{E, B_k\}=-i B_k~,\qquad\quad \{J_{kl},\,B_n\}=\delta_{kn}\,B_l-\delta_{ln}\,B_k~,\\[1mm]
		&\{B_k,\,B_l\}=0~,&    &\{B_k,\,B_l^*\}=2i\delta_{kl}\,E-2J_{kl}~.
	\ea\ee

	Based on \eqref{boosts 3} and \eqref{scalars1}, one can express $u_k$ and $v_k$ in terms of the boosts and perceive $B_k$ and $B_k^*$ as coordinates on the physical phase space.
	The two-form in \eqref{reduction 1} then becomes the symplectic form on the coadjoint orbit \cite{Dorn:2005jt},
	\be\ba\label{o=}
	\o=i\,\o_{kl}\,\,\dd B_k^* \wedge \dd B_l~, \qquad \text{with} ~~~~~~~~~~~~~~~~~~~~~~~~~~~~~\\[2mm]
	\o_{kl}=\frac{\d_{kl}}{2E}+\frac{(2E^2+|B|^2)B_k^* B_l -(2E^2-|B|^2)B_k B^*_l-
	{B^*}^2B_k B_l - B^2B_k^*B^*_l}{8E^3(2E^2-|B|^2-m^2)}~,
	\ea\ee
	where $|B|^2 \equiv B^*_n B_n$, ${B^*}^2 \equiv B^*_n B^*_n$ and $B^2 \equiv B_n B_n$, similarly to \eqref{scalarDefs}.
	Note that the matrix $\o_{kl}$ in \eqref{o=} is inverse to the matrix $\o^{kl} \equiv -i\{B_k, B_l^*\}$ given by \eqref{PB_Bn},
	which shows consistency of the Hamiltonian reduction.

	Now we construct a canonical parametrization of the physical phase space.
	The two-form in \eqref{reduction 1} contains a 'canonical part'  $m\,\dd u_k\wedge\dd  v_k$ and an additional
	two-form, which depends only on the scalar variables $\a$, $\b$ and $\g$. To compensate the latter and keep only the canonical structure,
	we are looking for the canonical variables in the form
	\be\label{Tr to CV 1}
		\begin{pmatrix} p_k\\ q_k \end{pmatrix}
			= \sqrt m \,\hat U \cdot \begin{pmatrix} u_k\\ v_k \end{pmatrix}~,
	\ee
	where $\hat U$ is a $2\times 2$ matrix with unit determinant and its components being $\mathrm{O}(N)$ scalars, {\it i.e.} depending only on $\a$, $\b$ and $\g$.
	With this ansatz one finds that the symplectic form in \eqref{reduction 1}
	reduces to $\o=\dd p_n\wedge\dd q_n$ for
	\be\label{hat M}
		\hat U=\f{1}{\sqrt{1+\a+\g+\d}} \begin{pmatrix}
			1+\a & \b\\
			\b & 1+\g
		\end{pmatrix}.
	\ee
	The scalar combinations of the canonical variables obtained from 
	\eqref{Tr to CV 1}-\eqref{hat M} read
	\be\label{scalars 1}
		p^2=m(\d+\g-\a-1)~, \qquad q^2=m(\d+\a-\g-1)~, \qquad p\cdot q=-2m\,\b~,
	\ee
	which allows to invert \eqref{Tr to CV 1} and  using then 
	\eqref{Noether charges}-\eqref{boosts 3}, 
	one finds a canonical parametrization
	of the Noether charges. In the oscillator variables,
	\be\label{b,b*}
		b_k=\f{p_k-i\,q_k}{\sqrt 2}~,  \qquad  b^*_k=\f{p_k+i\,q_k}{\sqrt 2}~,
	\ee
	this parametrization takes the following form
	\begin{align}\label{E=}
		\qquad&E=m+H~,& &J_{kl}=i(b_k^* b_l-b_l^* b_k)~,\\  \label{Bn=}
		&B_k= \left(Q +\f H {2Q}\right)\,b_k-\frac{b^2}{2Q}\,b^*_k~,&
			&B^*_k= \left(Q +\f H{2Q}\right)\,b^*_k-\frac{b^{*\,2}}{2Q}\,b_k~,
	\end{align}
	with
	\be\label{A=}
		H=b^*_n\,b_n~, \quad Q=\f 1 2\left(\sqrt{2m+H+J}+\sqrt{2m+H-J}\right)~,\quad J^2=H^2-b^{*\,2}b^2~.
	\ee

	From \eqref{E=}-\eqref{Bn=} and \eqref{scalars} follows that $B^2=\rho \,b^2$, with
	\be\label{F}
		\rho=\sqrt{(m+E)^2-J^2}=m(\a+\g)~,
	\ee
	which corresponds to the trace of the matrix \eqref{C}.
	This scalar function plays an important role in the geometric quantization of $\AdS$ particle \cite{Dorn:2005jt}.
	It characterizes the presymplectic form and the wave function of the ground state is just $\Psi_0=\rho^{-m}$.

	A similar canonical parametrization 
	of the $\SO(2,N)$ generators was considered in \cite{Dorn:2005ja} as a generalization of the Holstein-Primakoff representation to higher dimensions 
	(see Appendix \ref{app:bosAdS23}) and its validity was checked by the computation of the Poisson brackets
	algebra of the function \eqref{E=}-\eqref{Bn=}.

	The generators $E$ and $J_{kl}$ in \eqref{E=} are quadratic in canonical variables and their quantum realization is straightforward. However, the boost generators \eqref{Bn=} contain nontrivial operator ordering ambiguities. This problem was solved in \cite{Dorn:2005ja} using
	the results of geometric quantization \cite{Dorn:2005jt}, which lead to quantum deformations of the scalar functions involved in \eqref{Bn=}. 
	The exchange relations of these scalar functions 
	with the creation-annihilation operators then provide a quantum realization
	of \eqref{PB_Bn}.

\section{Dual Oscillators}\label{sec:DualOsc}

	Here, we propose a different quantization scheme based on dual oscillator variables.
	Using the notation $J=\sqrt{J^2}$, \eqref{Bn=} can be written as
	\be\label{B_k}
		B_k= {Q}\,b_k+ \tilde Q\,a_k~, 
	\ee\vspace{-1.5mm}
	with \vspace{-1.5mm}
	\begin{align}
		\label{Q}
		&\tilde Q=\f 1 2 \left(\sqrt{2m+H+J} - \sqrt{2m+H-J}\right)~,\\
		\label{a}
		&\qquad\quad a_k=\frac{i\,J_{kn}\,b_n}{J}=\f{H\,b_k-b^*_k\,b^2}{J}~,
	\end{align}
	and $k,n=1,\ldots,N$. With $a^*_k$ being the complex conjugated to \eqref{a}, one obtains the identities
	\be\ba\label{a to b}
		&a_n^*\, a_n=b_n^*\,b_n~,& \qquad &a_n a_n=-b_n b_n~,& \qquad  
		&i(a^*_ka_l-a^*_la_k)=i(b^*_k b_l-b^*_l b_k)~,& \\[2mm]
		&a_n\,b_n=0~,&    &a_n\,b_n^*=a_n^*\,b_n=J~.&    
	\ea\ee
	From  \eqref{a} then follows $b_k=i(J_{kn}\,a_n)/J$, where $J_{kn}$ and $J$ have the same form in terms of $a$ and $a^*$. 
	Moreover, the canonical relations 
	\be\label{canonical PB}
	\{b_k, b_l\}=0=\{b^*_k, b^*_l\}~, \qquad\qquad  \{b_k, b^*_l\}=i\d_{kl}~,
	\ee
	lead to the Poisson brackets
	\begin{align}\label{J a,b}
		&\{b_k, J\}=i\,a_k~,& &\{a_k, J\}=i\,b_k~,\\ \label{a,b}
		&\{a_k, b_l\}=\f i J\left(b^2 \d_{kl}+a_k a_l -b_k b_l\right),& \quad
		&\{a_k, b^*_l\}=\f i J\left(H \d_{kl}-a_k^* a_l -b^*_k b_l\right)\,,
	\end{align}
	and we find that $a_k$ and $a^*_k$ are also canonical,
	\be\label{canonical PB 1}
		\{a_k, a_l\}=0=\{a^*_k, a^*_l\}~,\qquad\qquad  \{a_k, a^*_l\}=i\d_{kl}~.
	\ee
	Hence, $a_k$ and $a^*_k$ are dual variables and the $\so(2,N)$ generators are symmetric under this duality. In the new variables
	\be\label{c-mp}
		c^{\pm}_k=\frac{b_k \pm  a_k}{\sqrt 2} ~, \qquad\qquad h_{\pm}=\f1 2(H \pm J)~,
	\ee
	the boost generators  \eqref{B_k} split into the sum
	\be\ba\label{Bk=}
	B_k=\sqrt{m+h_+}\,c^+_k+\sqrt{m+h_-}\,c^-_k\, , 
	\ea\ee
	and by \eqref{a,b}-\eqref{canonical PB 1} one obtains \vspace{-1.5mm}
	\be\ba\label{PB-pm1}
		&\{h_\pm, c_{k}^\mp\}=0~, &     &\{h_\pm, c_k^\pm \}=-i\,c_{k}^\pm~,&\\
		&\{c_{k}^\pm ,c_{l}^\pm \}=0~,&  
			&\{c_{k}^\pm ,c_{l}^\mp \}=\pm\frac{i}{J}\left(b^2\d_{kl}+a_k a_l-b_k b_l\right)~,&\\[-1mm]
		&\{c_{k}^\pm ,c^{\mp\,*}_{l}\}=0~,&  \qquad 
			&\{c_{k}^\pm ,c^{\pm\,*}_{l}\}= \pm\frac{i}{J}\left(2h_\pm\,\d_{kl}-a_k^* a_l-b^*_k b_l\right)~.&
	\ea\ee
	From \eqref{Bk=}-\eqref{PB-pm1} one easily checks the Poisson brackets
	\be\label{PB B,B^*}
		\{B_k, B_l\}=0=\{B^*_k, B^*_l\}~,  \qquad \{B_k, B_l^*\}=2i(m+H)\d_{kl}-2i(b^*_k b_l-b_l^* b_k)~,
	\ee
	which corresponds to the last two equations in \eqref{PB_Bn}. 
	Checking the remaining Poisson brackets of the $\so(2,N)$ algebra is obvious.   

	Now we discuss the quantum realization of the $\so(2,N)$ algebra in terms of dual oscillators. 
	The generators of  the compact subgroups are defined by
	\be\label{E-J op}
		E=m+H~, \qquad J_{kl}=i(b_k^\dag b_l-b_l^\dag b_k)~,
	\ee
	where $H = b^*_n\,b_n$ is the normal ordered harmonic oscillator Hamiltonian and $m$ corresponds to the minimal energy eigenvalue. 
	The angular momentum squared operator, {\it i.e.} the quadratic $\SO(N)$ Casimir, is defined in the usual way and one gets
	\be\label{J^2}
		J^2\equiv \f{1}{2}\, J_{ln}J_{ln}=H^2+(N-2)H-b^{*\,2}\,b^2~, 
	\ee
	which is an $N$-dependent quantum deformation of the classical relation in \eqref{A=}.

	We introduce the operator $J_N=\sqrt{J^2+\calC_N}$ as the quantum analog of the classical variable $J=\sqrt{J^2}$. Here, $\calC_N$ is an $N$-dependent number and on the basis of \eqref{J a,b} let us assume the commutation relations
	\be\label{b-a-J}
		[b_k, J_N]=a_k~, \qquad\qquad [a_k, J_N]=b_k~.
	\ee
	In Appendix \ref{app:DualOsc} we show that these conditions fix $\calC_N$ and $a_k$ by \cite{Heinze:footnote2} 
	\bea\label{C_N}
		&&\calC_N=\f 1 4(N-2)^2\\ \label{ak}
		&&a_k=\left(\Big(H+\f N 2\Big)b_k-b_k^*\, b^2\right)\f{1}{J_N}=
		\f{1}{J_N}\,\left(\Big(H+\f{N-2}{2}\Big)b_k-b_k^*\, b^2\right)~.
	\eea
	With this, one can then check that \eqref{b-a-J} is indeed fulfilled and
	the classical relations between the dual oscillator variables \eqref{a to b}
	become the following operator equations 
	\be\ba\label{a to b q}
		a_n^*\, a_n=b_n^*\,b_n~, \quad a_n a_n=-b_n b_n~, \quad  i(a^*_ka_l-a^*_la_k)=i(b^*_kb_l-b^*_lb_k)~,\\[2mm]
		\f 1 2(a_n\,b_n+b_n\,a_n)=0~, \quad   \f 1 2(a_n\,b_n^*+a_n^*\,b_n)=\f 1 2(b_n\,a_n^*+
		b_n^*\,a_n)=J_N~.   
	\ea\ee
	
	For example, to calculate $a_n^*\, a_n$ we use the first equation in \eqref{ak}, which yields
	\be\label{a*a=}
	a_n^*\, a_n=\f{1}{J_N}\,\left(b_n^*\Big(H+\f N 2\Big)-b^{*\,2}\,b_n\right) \left(\Big(H+\f N 2\Big)b_n-b_n^*\, b^2\right)\f{1}{J_N}~.
	\ee
	The product in the middle of the right hand side can be simplified and one obtains
	\be\label{a*a=1}
	a_n^*\, a_n=\f{1}{J_N} H \left(\Big(H+\f{N-2}{2}\Big)^2-b^{*\,2}\,b^2\right)\f{1}{J_N}=H~,
	\ee
	where we used
	\be\label{J_,H_N}
	J_N^2=H_N^2-b^{*\,2}b^2~, \qquad\qquad  \text{with} \qquad H_N=H+\f{N-2}{2}~.
	\ee
	The calculation of  $a_na_n$ involves both forms of $a_k$ in \eqref{ak}, which provides 
	\be\label{aa=}
	a_n\, a_n=\f{1}{J_N}\,\left(H_N\,b_n -b_n^*\,b^2\right) \left(\Big(H+\f N 2\Big)b_n-b_n^*\, b^2\right)\f{1}{J_N}~,
	\ee
	and the simplification of the operator expression here leads to
	$a_n a_n=-b_n b_n$.

	The check of the other operator in relations \eqref{a to b q} is similar, where one can use that the operator $J_N$ commutes with all $\mathrm{O}(N)$ scalar operators as well as with $J_{kl}$.

	Furthermore, from \eqref{ak} we obtain the canonical commutation relations  
	\be\label{canonical CR q}
		[a_k, a_l]=0=[a^*_k, a^*_l]~, \qquad\qquad  [a_k, a^*_l]=\d_{kl}~.
	\ee 
	The computation of the commutators between the dual oscillator operators is based on the operator identity \eqref{b_k:J} , which leads to the quantum version of \eqref{a,b}
	\bea\label{a,b q} 
		&&[a_k, b_l]=\Big(b^2 \d_{kl}+a_k a_l -b_k b_l\Big)\f 1 {J_N}=
		\f 1 {J_N}\Big(b^2 \d_{kl}+a_k a_l -b_k b_l\Big)~,\\[2mm] \label{a,b q1}
		&&[a_k, b^*_l]=\Big(H_N \, \d_{kl}-a_k^* a_l -b^*_k b_l\Big)\f 1{J_N}=
		\f 1{J_N}\Big( H_N\, \d_{kl}-a_k^* a_l -b^*_k b_l\Big)~. 
	\eea

	Similarly to \eqref{c-mp}, we define the operators
	\be\label{c-mp q}
		c^{\pm}_k=\frac{b_k \pm  a_k}{\sqrt 2} ~, \qquad\qquad h_{\pm}=\f1 2(H_N \pm J_N)~,
	\ee
	and with the help of \eqref{canonical CR q}-\eqref{a,b q1}, we obtain the quantum analog 
	of \eqref{PB-pm1} 
	\be\ba\label{Q-pm}
		&[h_\pm, c_{k}^\mp]=0~, &     &[h_\pm, c_k^\pm ]=-c_{k}^\pm~,&\\
		&[c_{k}^\pm ,c_{l}^\pm ]=0~,&  &[c_{k}^\pm ,c_{l}^\mp ]=\pm\frac{1}{J_N}\left(b^2\d_{kl}+
		a_k a_l-b_k b_l\right)~,&\\
		&[c_{k}^\pm ,c^{\mp\,*}_{l}]=0~,&  \qquad &[c_{k}^\pm ,c^{\pm\,*}_{l}]=
		\pm\frac{1}{J_N}\left(2h_\pm\,\d_{kl}-a_k^* a_l-b^*_k b_l\right)~.&
	\ea\ee

	Finally, using the classical expression \eqref{Bk=}, we introduce the boost operators by 
	\be\ba\label{q Bk=}
		B_k=\sqrt{m_N+h_+}\,c^+_k+\sqrt{m_N+h_-}\,c^-_k\,, 
		\quad B_k^*=c^{+\,*}_{k}\sqrt{m_N+h_+}+c^{-\,*}_{k}\sqrt{m_N+h_-}\,,
	\ea\ee
	where $m_N$ is a deformed mass parameter. 
	Equations \eqref{Q-pm} lead to the commutators
	\be\label{Comm B,B^*}
	[B_k, B_l]=0=[B^*_k, B^*_l]~,  \qquad [B_k, B_l^*]=2(m_N+H_N)\d_{kl}-2(b^*_k b_l-b_l^* b_k)~,
	\ee
	which becomes a part of the $\so(2,N)$ algebra for $m_N=m-\f{N-2}{2}$.
	The representation \eqref{q Bk=} is unitary if the square root expressions are Hermitian. This requires $m\geq\f{N-2}{2}$, which coincides with
	the unitarity bound of the $\so(2,N)$ representations \cite{Breitenlohner:1982jf}.

\section{Supersymmetric Extension} \label{sec:Super}
	In this section we discuss a supersymmetric extension of the $\so(2,N)$ algebra. Furthermore, we will show how the $\SO(N)$ covariant oscillator representation can be supersymmetrized, generalizing the known results for the $\calN=1$ $\AdS_2$ \cite{Heinze:2015oha} and $\AdS_3$ \cite{Heinze:2016fin} superparticles.


	For this, let $\G_A$ be the $\SO(2,N)$ gamma matrices forming the Clifford algebra 
	\be\label{Clifford algebra}
		\G_A \G_B + \G_B \G_A=\eta_{AB} I~,\qquad\qquad A,B=0',0,1,\ldots,N~.
	\ee
	Their normalized commutators $\G_{AB}=\f 1 4[\G_A, \G_B]$ provide the $\so(2,N)$ algebra \eqref{so(2,N) matrix algebra},
	\be\label{so(2,N) algebra 1}
		[\G_{AB}, \G_{CD}]=\eta_{AD}\G_{BC}+\eta_{BC}\G_{AD}-\eta_{AC} \G_{BD}-\eta_{BD}\G_{AC}~.
	\ee
	The charge conjugation matrix $C_\epsilon$ is introduced by its defining property
	\be\label{C-matrix}
		\G_A^T=\epsilon\, C_\epsilon \,\G_A\, C_\epsilon^{-1}~,
		\qquad\qquad \text{with} \quad \epsilon=\pm 1~,
	\ee
	where we will assume that $C_\epsilon$ is antisymmetric.
	The matrices $\G_{AB} C_\epsilon^{-1}$ are then symmetric and
	\be\label{G-C}
		C_\epsilon^{-1} \G_{AB}^T=-\G_{AB} C_\epsilon^{-1}~.
	\ee
	Note that $C_+$ exists for $N=2, 3, 4$ modulo 8 while $C_-$ exists for $N=0, 1, 2$ modulo $8$.

	For the bosonic charges $J_{AB}$ fulfilling the $\so(2,N)$ Poisson algebra \eqref{PB_J}, let us introduce its supersymmetric extension by the following additional commutation relations 
	\be\label{superalgebra}
		\{J_{AB}, F_a\}=\left(\G_{AB}\right)_{ab} F_b~, \qquad \{F_a, F_b\}=\left(\G^{AB} C_\epsilon^{-1}\right)_{ab} J_{AB}~,
	\ee
	where $F_a$ are the odd elements of the algebra forming an $\SO(2,N)$ spinor.

	The Jacobi  identity with two even and one odd elements
	\be\label{jacobi 2-1}
		\{J_{AB},\{J_{CD}, F_a\}\}+\{J_{CD},\{F_a, J_{AB}\}\}+\{F_a,\{ J_{AB}, J_{CD}\}\}=0~,
	\ee
	trivially follows from \eqref{PB_J}, \eqref{so(2,N) algebra 1} and \eqref{superalgebra}. The case of
	one even and two odd elements
	\be\label{jacobi 1-2}
		\{J_{AB},\{F_a, F_b\}\}+\{F_a,\{F_b, J_{AB}\}\}-\{F_b,\{J_{AB}, F_a\}\} = 0~,
	\ee
	is satisfied because of \eqref{G-C} and the antisymmetricity of the structure constants.  
	Finally, the case of three odd elements,
	\be\label{jacobi 0-3}
		\{F_c,\{F_a, F_b\}\}+\{F_a,\{F_b, F_c\}\}+\{F_b,\{F_c, F_a\}\} =0~,
	\ee
	holds if the gamma matrices satisfy the relation
	\be\label{jacobi 3}
		\left(\G^{AB} C_\epsilon^{-1}\right)_{ab} \left(\G_{AB}\right)_{cd}+\left(\G^{AB} 
		C_\epsilon^{-1}\right)_{bc} \left(\G_{AB}\right)_{ad}+
		\left(\G^{AB} C_\epsilon^{-1}\right)_{ca} \left(\G_{AB}\right)_{bd} = 0~.
	\ee

	It is well known that superconformal algebras can be found only for $N \leq 6$. Requiring existence of the charge conjugation matrix $C_\eps$ reduces the possible values of $N$, for which the above relation might hold, even further to $N\leq4$. 
	In fact, we have checked directly (for $N\leq10$)  that \eqref{jacobi 3} can be satisfied only for $N\leq3$. This observation is in accordance with the fact that only for $N\leq3$ there exist superconformal algebras without any $R$-symmetry, as is the case for \eqref{superalgebra}.

	For $N=1$, {\it i.e.} $\AdS_2$, we can realize $\G_A$'s and $C_\eps$ by the Pauli matrices
	\be\label{pauli m}
		\G_{0'}=i\s_1~, \qquad \G_0=i\s_3~, \qquad \G_1=\s_2~, \qquad C_{-}=\s_2~,
	\ee
	and \eqref{jacobi 3} reduces to the identity
	\be\label{Pauli identity}
	\d_{ab} (\s_2)_{cd}=i\big((\s_1)_{ac}(\s_3)_{bd}-(\s_1)_{ad}(\s_3)_{bc}\big)~.
	\ee

	The obtained superalgebra is $\osp(1|2)$ and 
	the $\OSP(1|2)/\SO(1,1)$ coset scheme leads to the Holstein-Primakoff type 
	representation of 
	$\osp(1|2)$  \cite{Heinze:2015oha}.  

	Using canonical fermionic creation-annihilation variables $f$ and $f^*$, with $\{f, f^*\}=i$, the parametrization of the $\osp(1|2)$ generators can be written as 
	\be\label{b,f to B,F}
		E=m+b^* b+\frac{f^* f}{2}~,\quad B=\sqrt{2m+b^* b+f^* f}\,b~, \quad  
		F=\sqrt{2m+b^* b+f^* f}\,f+f^*\,b~.
	\ee
	which is a supersymmetric extension of \eqref{B=b}.

	The case of $N=2$ is given by the $\AdS_3$ superparticle on $\OSP(1|2)\times\OSP(1|2)/\SLS(2,\Reals)$ \cite{Heinze:2016fin}. Here, the superisometry algebra consists of the direct sum $\osp_l(1|2) \oplus \osp_r(1|2)$, with each $\osp(1|2)$ \eqref{b,f to B,F} given in terms of 'left' and 'right' canonical variables 
	\be\ba\label{b,f to B,F AdS3}
	&E_l=\f{m}{2}+b_l^* b_l+\frac{f_l^* f_l}{2}\,,& \qquad   &E_r=\f{m}{2}+b_r^* b_r+\frac{f_r^* f_r}{2}\,, &  \\
	&B_l=\sqrt{m+b_l^* b_l+f_l^* f_l}\,\,b_l\,,&             &B_r=\sqrt{m+b_r^* b_r+f_r^* f_r}\,\,b_r\,,&\\
	&F_l=\sqrt{m+b_l^* b_l}\,f_l+f_l^*\,b_l\,,&        &F_r=\sqrt{m+b_r^* b_r}\,f_r+f_r^*\,b_r\,.& 
	\ea\ee
	The following linear combinations of the bosonic elements
	\be\label{AdS_3 generators}
	E=E_l+E_r~, \quad  J_{12}=E_r-E_l~, \quad B_1=B_l+B_r~, \quad B_2=i(B_l-B_r)~,
	\ee
	provide the $\so(2,2)$ algebra written in the general $\SO(N)$ covariant form \eqref{PB_Bn},
	where the indices take the values $k,l,m=1,2$.

	We furthermore introduce the $\SO(2)$ spinor $F_\a$, $\a=1,2$, and its complex conjugate $F_\a^*$ by 
	\be\label{AdS_3 fermionic}
		F_1 = F_l+F_r~, \qquad\qquad F_2 =i(F_l-F_r)~,
	\ee
	as well as the gamma matrices $\g_1=\sigma_3$ and $\g_2=\sigma_1$. Then, the remaining $\osp_l(1|2)\oplus \osp_r(1|2)$ Poisson brackets take the $\SO(N)$ covariant form
	\be\ba\label{algebra with fermions}
		&\{E, F_\a\}=-i\,F_\a/2~,&  \quad  &\{J_{mn}, F_\a\}=\left(\g_{mn}\right)_{\a\b}F_\b~,& \\
		&\{B_n, F_\a\}=0~,& &\{B^*_n , F_\a\}=-i(\g_n)_{\a\b}F^*_\b~,&\\
		&\{F_\a , F_\b \}=2i(\g_n)_{\a\b} B_n~,&  &\{F_\a^* , F_\b\}=2iE\d_{\a\b}+2J_{mn}(\g_{mn})_{\a\b}~,& 
	\ea\ee
	with $\g_{mn}=\f{1}{4}[\g_m,\g_n]$, $m,n=1,2,\ldots,N$ and $N=2$ for the case at hand. 

	To show that the above is equivalent to the $\SO(2,N)$ covariant algebra \eqref{superalgebra} we take for the $\SO(2,2)$ spinor $F_a$, $a=1,2,3,4$, the natural ansatz $F_a = (F_\alpha,F^*_\alpha)$, which is related to the manifestly real Majorana spinor by the similarity transformation
	\be
		\begin{pmatrix} \frac{1}{\sqrt{2}}(F_\a + F^*_\a)\\ \frac{i}{\sqrt{2}}(F_\a - F^*_\a)\end{pmatrix} = 
			\frac{1}{\sqrt{2}} \left(\begin{pmatrix} 1&1\\i&-i\end{pmatrix}\times \mathbbm{1}_{2\times2}\right)
			\begin{pmatrix} F_\a \\ F^*_\a \end{pmatrix}~.
	\ee
	
	Recalling that $E=J_{0'0}$ and $B_m =J_{m 0'}-i\,J_{m 0}$, \eqref{algebra with fermions} then matches with \eqref{superalgebra} if for the $\SO(2,2)$ gamma matrices we take the basis
	\be
		\G_{0'} = -i\,\s_1 \times \mathbbm{1}_{2\times2}~,\qquad
		\G_{0} = -i\,\s_2 \times \mathbbm{1}_{2\times2}~,\qquad
		\G_{m=1,2} = \s_3 \times \g_m~
	\ee
	and for the charge conjugation matrix
	\be
		C_\eps = C_- = \frac{1}{2}\,\G_0 = \frac{-i}{2}\,\s_2 \times \mathbbm{1}_{2\times2}~,
	\ee
	where the prefactor of $1/2$ could also have been absorbed in the definition of $F_a$.

	With the superalgebra in an $\SO(N)$, respectively, $\SO(2,N)$ covariant form we recall that in the bosonic case also the oscillator parametrization of the charges \eqref{E=}-\eqref{Bn=} became covariant under the spatial $\SO(N)$. Indeed, in the new oscillator variables
	\be\label{b-f}
		b_1=\f{b_l+b_r}{\sqrt 2}~, \quad  b_2=i\f{b_l-b_r}{\sqrt 2}~, \qquad
		f_1=\f{f_l+f_r}{\sqrt 2}~, \quad  f_2=i\f{f_l-f_r}{\sqrt 2}~,
	\ee
	also the generators \eqref{AdS_3 generators}-\eqref{AdS_3 fermionic} take an $\SO(N)$ covariant form,
	\be\ba\label{symmetry generators}
		&E=m+H+\frac{h}{2}~, \quad  B_m=Q\, b_m+\f{i}{2Q}\,({J}_{mn}+j_{mn})b_n~,\quad  
		J_{mn}=\tilde{J}_{mn}+j_{mn}~,\\
		&F_\a=Q \,f_a+\f{i}{2Q}\,({J}_{mn}+j_{mn})\left(\g_{mn}\right)_{\a\b}f_\b+f_\b^*(\g_n)_{\b\a}b_n~,
	\ea\ee
	with
	\be\ba
		&H=b_m^*b_m~,\quad h=f_\b^*f_\b~, \quad  \tilde{J}_{mn}=i(b_m^* b_n-b^*_m b_n)~, 
		\quad j_{mn}=i\,f_\a^*\left(\g_{mn}\right)_{\a\b}f_\b~,\\
		&Q=\f{1}{2}\left(\sqrt{2m+H+h+(J+j)}+\sqrt{2m+H+h-(J+j)}\right)~,\\
		&({J}+j)^2=\f{1}{2}({J}_{mn}+j_{mn})({J}_{mn}+j_{mn})~.
	\ea\ee
	Hence, these should be viewed as a supersymmetric generalization of \eqref{E=}-\eqref{Bn=}.
	Although \eqref{symmetry generators} hold for $N=1$ and $N=2$, showing consistency for $N=3$, which ought to correspond to the $\calN=1$ $\AdS_4$ superparticle, is still an open problem. For $N\geq4$ the ansatz for the charges \eqref{symmetry generators} has to fail due to inconstancy of \eqref{jacobi 3}, as stated above.


\subsection*{Acknowledgments}
\noindent 
We thank Gleb Arutyunov, Harald Dorn, Tomas Klose and Jan Plefka for useful discussions.
M.H. thanks the organizers of the conference {\it Selected Topics in Theoretical High Energy Physics} (Tbilisi, 2015) 
and of the program {\it Holography and Dualities 2016: New Advances in String and Gauge Theory} (Nordita, 2016) as well as Nordita in Stockholm for kind hospitality.
G.J. thanks the Humboldt University of Berlin and University of Hamburg
for kind hospitality.
The work  M.H. and G.J. is  supported  by the German Science Foundation (DFG) under the
Collaborative Research Center (SFB) 676 {\it Particles, Strings and the Early Universe}.
The work of G.J. and L.M. is supported by the Rustaveli GNSF. In addition, G.J. is supported by the DFG under the SFB 647 {\it Space--Time--Matter}.



\appendix

\section{Particle Dynamics in \texorpdfstring{$\AdS_{N+1}$}{AdSN+1}} \label{app:AdSNpl1}

	Here we consider a standard formulation of particle dynamics in $\AdS_{N+1}$ and discuss its relation to
	the $\SO(2,N)/\SO(1,N)$ coset construction.

	Let us represent $\AdS_{N+1}$ as the hyperboloid of a radius $R$ embedded in $\Reals^{2,N}$
	\be\label{hyperbola}
	X^A X_A+R^2=0~,
	\ee
	and introduce the action of AdS particle by
	\begin{equation}\label{S}
	S=\int d\tau \left(\frac{\dot X^A\dot X_A}{2\xi}-\frac{\xi M^2}{2}\right)~.
	\end{equation}
	Here, $M$ is the particle mass, $\xi$ plays the role of einbein and $X^A$ satisfies  \eqref{hyperbola}.
	This action reduces to the action of the coset model \eqref{AdS coset action 1-2} by 
	the following rescaling of variables
	\be\label{rescaling}
	x^A=X^A/R~,\qquad  e=\xi/R^2~,  \qquad m=MR~.
	\ee

	The $\SO(2,N)$ symmetry of \eqref{hyperbola}-\eqref{S} defines the Noether charges
	\begin{equation}\label{M_AB}
	J_{AB} =p_A\,x_B -p_B\,x_A~,
	\end{equation}
	where $p_A= {\dot x_A}/e$  are the canonical momenta and the canonical Poisson brackets
	$\{p_A,\,x^B\}=\d_A^{\,\,B}$ provide the $\so(2,N)$ algebra \eqref{PB_J} for these charges.

	In the first order formalism Dirac's procedure leads to the constraints
	\begin{equation}\label{Phi=0}
	x^A x_A+1=0~,\qquad\qquad p_A p^A+m^2=0~,\qquad\qquad p_A\,x^A=0~.
	\end{equation}
	The linear combination $\Phi_0=m^2(x^A x_A+1)+(p_A p^A+m^2)$
	is a first class constraint and two others, $\Phi_1=p_A\,x^A$ and $\Phi_2=x^A x_A+1$, are of the second class.
	Imposing then the gauge fixing condition $m\,x_0+p_{0'}=0$, we find that the description in the phase space variables $p_A$ and $x^A$
	is equivalent to \eqref{reduction 1}, with $p_A=m\,u_A$ and $x^A=v^A$.

	The constraints \eqref{Phi=0} provide the mass-shell condition \eqref{mass-shell}
	and from \eqref{M_AB} also follow the quadratic relations 
	between the Noether charges,
	\begin{equation}\label{MM=MM}
	J_{AB}\,J_{CD}+ J_{AC}\,J_{DB}+J_{AD}\,J_{BC}=0~.
	\end{equation}
	Taking $A=0'$, $B=0$, $C=k$ and $D=l$, together with \eqref{B,B*} this yields the identity
	\begin{equation}\label{EM=zz}
	2E\,J_{kl}=i(B_k^*B_l-B_l^*B_k)~,
	\end{equation}
	and by the mass-shell condition \eqref{mass-shell} and \eqref{EM=zz} one obtains two $\mathrm{O}(N)$ scalar relations
	\be\label{E^2=}
	E^2+J^2=m^2+|B|^2~, \qquad\qquad 4E^2J^2=|B|^4-{B^*}^2B^2~,
	\ee
	where the scalar combinations are defined as in \eqref{o=}.
	From \eqref{EM=zz}-\eqref{E^2=} one then finds the generators $E$ and $J_{kl}$ as functions of $B_k$ and $B_k^*$.
	As a result, $B_k$ and $B_k^*$ become global coordinates on the physical phase space. 

	Finally, note that the constraints \eqref{Phi=0} provide the relation
	\begin{equation}\label{omega0}
	\frac{1}{2m^2}J_{AB}\,dJ^{AC}\wedge
	\,dJ^B\,_C=dP_A\wedge dX^A~,
	\end{equation}
	which parameterizes the symplectic form by the Noether charges and leads to \eqref{o=}.

\section{Parametrization of \texorpdfstring{$\SO(2,N)$}{SO2N} Boosts} \label{app:CosetParam}

	In this appendix we analyze the structure of the boost matrix $g_b$ defined by \eqref{decompositiom}
	and show that the matrix elements of $C$ in \eqref{C} are uniquely defined by  \eqref{constraints 4}.

	The matrix $g_b$ is an exponent $g_b=e^T$, where $T\in \so(2,N)$ has the  block structure
		\be\label{boost g-1} 
			T=\begin{pmatrix}
			& \zeta^\a_{~\,\,l}\\
			\zeta^k_{~\,\,\b} & 
		\end{pmatrix}.
	\ee
	Here, the indices take the values $\a, \b = 0',0$ and $k=1,2,\dots, N$ and we left the vanishing upper-left $2\times2$ block and lower-right $N \times N$ block blank. Thus, $\zeta^\a_{~\,\,l}$ and $\zeta^k_{~\,\,\b}$
	are $2\times N$ and $N\times 2$ matrices, respectively, and by \eqref{boost g-1} one finds
		\be\label{boost g-2} 
			T^2=\begin{pmatrix}
			\xi^\a_{~\,\,\b}  & \\
			& \zeta^k_{~\,\,\g}\,\zeta^\g_{~\,\,l}
			\end{pmatrix},
		\ee
	with 
	\be\label{xi}
	\xi^\a_{~\,\,\b} =\zeta^\a_{~\,\,j}\,\zeta^j_{~\,\,\b}=\zeta^{\a j}\,\zeta^{\b j}~. 
	\ee

	Using \eqref{boost g-1}-\eqref{boost g-2} and recurrence relations, one finds
	\be\label{boost g-1-2}
		T^{2n}=\begin{pmatrix}
		(\xi^n)^\a_{~\,\,\b}  & \\
		 & \zeta^k_{~\,\,\g}\,(\xi^{n-1})^\g_{~\,\,\g'}\zeta^{\g'}_{~\,\,l}
		\end{pmatrix},\qquad
		T^{2n+1}=\begin{pmatrix}
		 & (\xi^n)^\a_{~\,\,\g}\,\zeta^\g_{~\,\,l}\\
		\zeta^k_{~\,\,\g}\,(\xi^n)^\g_{~\,\,\b} & 
		\end{pmatrix}.
	\ee
	The calculation of the exponent $e^T$ is then straightforward and one obtains
	\be\label{boost} 
		g_b=\begin{pmatrix}
		C^\a_{~\,\,\b}  & S^\a_{~\,\,\g}\,\zeta^\g_{~\,\,l}\\
		\zeta^k_{~\,\,\g}\,S^\g_{~\,\,\b}  & \zeta^k_{~\,\,\g}\,U^\g_{~\,\,\g'}
		\zeta^{\g'}_{~\,\,l}
		\end{pmatrix},
	\ee
	where
	\be\label{sums}
		C^\a_{~\,\,\b}=\sum_{n=0}^{\infty} \f{(\xi^n)^\a_{~\,\,\b}}{(2n)!}~, \qquad
		S^\a_{~\,\,\b}=\sum_{n=0}^{\infty} \f{(\xi^n)^\a_{~\,\,\b}}{(2n+1)!}~, \qquad
		U^\a_{~\,\,\b}=\sum_{n=0}^{\infty} \f{(\xi^{n-1})^\a_{~\,\,\b}}{(2n)!}~.
	\ee

The matrix $\xi^\a_{~\,\,\b}$ given by \eqref{xi} is symmetric and semi-positive.
Therefore, it can be represented as a square of a symmetric and semi-positive matrix $\xi=\eta^2$ and  one gets $C=\cosh \eta$. As a result, the matrix $C$ is also symmetric and positive. Equations \eqref{constraints 4} then uniquely define the matrix elements $\a$, $\b$ and $\g$ 
and one obtains 
\be\label{abc}
	\a=\f{1+u^2+X}{\sqrt{2+u^2+v^2+2X}}\,,\quad 
		\b=\f{u v}{\sqrt{2+u^2+v^2+2X}}\,, \quad
		\g=\f{1+v^2+X}{\sqrt{2+u^2+v^2+2X}}\,,
\ee
where $X=\sqrt{1+u^2+v^2+u^2v^2-(u\,v)^2}$. 

Note that there is a one-to-one correspondence between the group parameters 
$\zeta^{0'k}$ and $\zeta^{0k})$ and the variables $u_k$ and $v_k$. At $\zeta^{0'k}=0$ equations \eqref{boost}-\eqref{abc} reproduce the more familiar structure of the $\SO(1,N)$ boosts.

\section{Canonical Parametrization of \texorpdfstring{$\SO(2,N)$}{SO2N} Generators}\label{app:bosAdS23}

Here we analyze the canonical structure of AdS particle dynamics in low dimensions and show
that its generalization leads to the canonical parameterization \eqref{Bn=}.

$\AdS_2$ corresponds to $N=1$. In this case \eqref{mass-shell} and \eqref{o=}
are equivalent to
\be\label{AdS_2}
  E^2=m^2+B^*B~,  \qquad\qquad     \o=i\,\frac{\dd B^*\wedge \dd B}{2E}~,
\ee
with $B=B_1,$ $B^*=B_1^*$. The Holstein-Primakoff parametrization
\be\label{B=b}
E=m+b^* b~, \qquad  B= \sqrt{2m+b^* b}\,\,b~, \qquad B^*=b^*\,\sqrt{2m+b^* b}
\ee
then reduces the symplectic form in \eqref{AdS_2} to the canonical one $\o=i\, \dd b^*\wedge \dd b$.

	In $\AdS_3$ equations \eqref{EM=zz} and the mass-shell condition provide the relations
	\begin{equation}\label{2eq}
		2E\,J_{12}=i(B^*_1B_2-B^*_2B_1)~, \qquad  E^2+J_{12}^2= m^2+|B_1|^2 + |B_2|^2~,
	\end{equation}
	which are equivalent to
	\begin{equation}\label{C_2}
		E^2_L=\left({m}/{2}\right)^2+|B_L|^2~,\qquad\qquad
		E^2_R=\left({m}/{2}\right)^2+|B_R|^2~,
	\end{equation}
	where the `left' ($E_L, B_L, B_L^*$) and the `right' ($E_R, B_R, B_R^*$) variables  are defined by
	\be\label{L-R}
	2E_L=E-J_{12}~,\quad 2B_L=B_1-i B_2~, \qquad 2E_R=E+J_{12}~, \quad 2B_R=B_1+i B_2~.
	\ee
	By \eqref{PB_J}, the left variables $E_L$, $B_L$, and $B_L^*$ have
	zero Poisson brackets with the right variables $E_R$, $B_R$, and $B_R^*$ and they both
	form an $\so(2,1)$ algebra, as $E$, $B$ and $B^*$ do for $N=1$.

	Taking into account that the Casimir parameter of the left and
	right parts is $m/2$, we introduce the parametrization
	\eqref{B=b} for the left and right variables as follows
	\begin{equation}\label{L-R_a}
		E_L=\f{m}{2}+H_L\,,\quad B_L=\sqrt{m+H_L}\,\,b_L\,, \qquad E_R=\f{m}{2}+H_R\,,\quad B_R=\sqrt{m+H_R}\,\,b_R\,,
	\end{equation}
	with $H_L=b_L^*b_L$ and $H_R=b_R^*b_R$. From \eqref{C_2} and \eqref{L-R} we then find
	\be\ba\label{M_12}
		&E=H_L+H_R+m&   &J_{12}=H_R-H_L &\\
		&B_1=\sqrt{m+H_L}\,\,b_L+\sqrt{m+H_R}\,\,b_R &   &B_2=i\sqrt{m+H_L}\,\,b_L-i \sqrt{m+H_R}\,\,b_R~.&\\
	\ea\ee

	The obtained canonical form of the $\SO(2,2)$ generators is not
	suitable for a generalization to arbitrary $N$ and to proceed,
	we introduce the new canonical variables
	\begin{eqnarray}\label{a_1,2}
		b_1=\frac{b_L+b_R}{\sqrt{2}}~,\quad b^*_1=\frac{b^*_L+b^*_R}{\sqrt{2}}~; \quad
		b_2=i\,\frac{b_L-b_R}{\sqrt{2}}~,\quad b^*_2=i\,\frac{b^*_R-b^*_L}{\sqrt{2}}~.
	\end{eqnarray}
	This provides $H_R-H_L=i(b_1^*b_2-b_2^*b_1)$, $H_R+H_L=|b_1|^2+|b_2|^2$ and
	the $SO(2,2)$ generators \eqref{M_12} become
	\be\ba\label{E,M_12=a1a2}
		&E=m+H~, \qquad  ~~~~~~~~~~ J_{12}=i(b_1^*b_2-b_2^*b_1)~,\\
		&B_1=Q \,b_1+\f{i}{2\,Q}\,{J_{12} b_2}~, \quad
		B_2=Q\,b_2+\f{i}{2\,Q}\,{J_{21} b_1}~,~~~\mbox{with} \\
		&Q=\f 1 2\left(\sqrt{2m+H+J}+\sqrt{2m+H-J}\right)~,~~  H=|b_1|^2+|b_2|^2~,~~  J=\sqrt{J_{12}^2}~.
	\ea\ee
	This equation defines a parametrization
	of the symmetry generators in terms of canonical variables, which has a natural generalization
	for arbitrary $N$ in the form \eqref{Bn=}.

\section{More on Dual Oscillators} \label{app:DualOsc}

	In this appendix we present calculations related to quantized dual oscillators.

	First we fix the operators $J_N$ and $a_k$ on the basis of \eqref{b-a-J},
	which provides 
	\be\label{bk J^2}
	[b_k, J_N^2]=2\,a_k\,J_N-b_k, \qquad\qquad [a_k, J_N^2]=2\,b_k\,J_N-a_k~.
	\ee
	On the other hand, the form of the operaor $J_N$ together with \eqref{J^2} yields 
	\be\label{b-J^2}
	[b_k, J_N^2]=(2H+N-1)b_k-2b_k^*\,b^2~,
	\ee
	with $b^2 = b_n\,b_n$ and ${b^*}^2 = b^*_n\,b^*_n$. Comparing this with the first equation of \eqref{bk J^2}, we find
	\be\label{ak J}
	a_k\,J_N=(H+N/2)b_k-b^*_k\,b^2~.
	\ee
	From this follows the commutator
	\be\label{akJ-J^2}
	[a_k\,J_N, J_N^2]=(H+N/2)\Big((2H+N-1)b_k-2b^*_k\,b^2\Big)+\Big(b_k^*(2H+N-1)-2b^{*\,2}b_k\Big)b^2,
	\ee
	and by the second equation of \eqref{bk J^2} we have $[a_k\,J_N, J_N^2]=2b_kJ_N^2-a_k\,J_N$. Inserting here \eqref{b-J^2}-\eqref{ak J} and comparing it with \eqref{akJ-J^2} fixes the number $\calC_N$  by \eqref{C_N}. The operator $J_N$ has then an inverse  and $a_k$ is given  by the first equation of \eqref{ak}.

	The obtained $a_k$ and $J_N$ satisfy the exchange relations
	\be\label{exch relations a-b} 
	J_N^2 \,b_k=b_k\,J_N^2-2a_k\,J_N+b_k~, \qquad J_N^2 \, a_k=a_k\,J_N^2-2b_k\, J_N+a_k~,
	\ee
	which are equivalent to 
	\be\label{exch relations c} 
	J_N^2 \,c^+_k=c^+_k\,(J_N-1)^2~, \qquad\qquad J_N^2 \, c^-_k=c^-_k (J_N+1)^2~,
	\ee
	with $c_k^\pm$ defined by \eqref{c-mp}. One then gets $J_N \,c^\pm_k=c^\pm_k\,(J_N\mp 1)$
	and we indeed obtain \eqref{b-a-J}.

	The first equation of \eqref{bk J^2} can also be written as $[b_k, J_N^2]=2\,J_N\,a_k-b_k$,
	which together with \eqref{b-J^2} provides the second equation of \eqref{ak}.
	These two forms of $a_k$ allow to check the operator relations \eqref{a to b q}
	and the commutators \eqref{a,b q}-\eqref{canonical CR q}. At some points one can 
	also use the equation
	\be\label{b_k:J}
	\f 1{J_N}\,b_l=b_l\,\f 1{J_N}+\f 1{J_N}\,a_l\,\f 1{J_N}~,
	\ee
	which follows from \eqref{b-a-J}. For example, calculating $a_k b_l$, one finds
	\be\label{ab=}
	a_k b_l=a_k J_N \f 1{J_N} b_l=a_k J_N \left(b_l \f 1{J_N}+\f 1{J_N} a_l \f 1{J_N}\right)=
	b_l a_k +\left([a_k J_N, b_l]+a_k\,a_l\right)\f 1{J_N}~,
	\ee
	and from \eqref{ak J} follows the commutator  $[a_k J_N, b_l]=b^2\d_{kl}-b_kb_l$.
	Equation \eqref{ab=} is then equivalent to the first form of the commutator \eqref{a,b q}.
	The second form is obtained in a similar way, applying \eqref{b_k:J} to the term $b_l a_k$.
	Two forms of the commutator $[a_k, b^*_l]$ is obtained in the same way and they finally lead to the commutators of the boosts \eqref{Comm B,B^*}.
	


\bibliographystyle{../../../../../Latex/bibSpires/nb}
\bibliography{../../../../../Latex/bibSpires/bibSpires}

\begin{thebibliography}{10}
\ifx\href\asklfhas\newcommand{\href}[2]{#2}\fi
\ifx\arxivref\asklfhas\newcommand{\arxivref}[2]{\href{http://arxiv.org/abs/#1}{#2}}\fi
\ifx\doiref\asklfhas\newcommand{\doiref}[2]{\href{http://dx.doi.org/#1}{#2}}\fi
\raggedright
\small
\parskip 0pt

\bibitem{Arutyunov:2009ga}
G.~Arutyunov and S.~Frolov,
\textit{``{Foundations of the $AdS_5 \times S^5$ Superstring. Part I}''},
\textsf{\doiref{10.1088/1751-8113/42/25/254003}{J.~Phys.~A42,~254003~(2009)}},
\texttt{\arxivref{0901.4937}{arxiv:0901.4937}}.

\bibitem{Beisert:2010jr}
N.~Beisert, C.~Ahn, L.~F.~Alday, Z.~Bajnok, J.~M.~Drummond et~al.,
\textit{``{Review of AdS/CFT Integrability: An Overview}''},
\textsf{\doiref{10.1007/s11005-011-0529-2}{Lett.Math.Phys.~99,~3~(2012)}},
\texttt{\arxivref{1012.3982}{arxiv:1012.3982}}.

\bibitem{Bombardelli:2016rwb}
D.~Bombardelli, A.~Cagnazzo, R.~Frassek, F.~Levkovich-Maslyuk, F.~Loebbert,
  S.~Negro, I.~M.~Szécsényi, A.~Sfondrini, S.~J.~van~Tongeren and
  A.~Torrielli,
\textit{``{An integrability primer for the gauge-gravity correspondence: An
  introduction}''},
\textsf{\doiref{10.1088/1751-8113/49/32/320301}{J.~Phys.~A49,~320301~(2016)}},
\texttt{\arxivref{1606.02945}{arxiv:1606.02945}}.

\bibitem{Arutyunov:2009ur}
G.~Arutyunov and S.~Frolov,
\textit{``{Thermodynamic Bethe Ansatz for the $AdS_5\times S^5$ Mirror
  Model}''},
\textsf{\doiref{10.1088/1126-6708/2009/05/068}{JHEP~0905,~068~(2009)}},
\texttt{\arxivref{0903.0141}{arxiv:0903.0141}}.

\bibitem{Bombardelli:2009ns}
D.~Bombardelli, D.~Fioravanti and R.~Tateo,
\textit{``{Thermodynamic Bethe Ansatz for planar AdS/CFT: a proposal}''},
\textsf{\doiref{10.1088/1751-8113/42/37/375401}{J.~Phys.~A42,~375401~(2009)}},
\texttt{\arxivref{0902.3930}{arxiv:0902.3930}}.

\bibitem{Gromov:2009tv}
N.~Gromov, V.~Kazakov and P.~Vieira,
\textit{``{Exact Spectrum of Anomalous Dimensions of Planar N=4 Supersymmetric
  Yang-Mills Theory}''},
\textsf{\doiref{10.1103/PhysRevLett.103.131601}{Phys.Rev.Lett.~103,~131601~(2009)}},
\texttt{\arxivref{0901.3753}{arxiv:0901.3753}}.

\bibitem{Gromov:2009bc}
N.~Gromov, V.~Kazakov, A.~Kozak and P.~Vieira,
\textit{``{Exact Spectrum of Anomalous Dimensions of Planar N = 4
  Supersymmetric Yang-Mills Theory: TBA and excited states}''},
\textsf{\doiref{10.1007/s11005-010-0374-8}{Lett.~Math.~Phys.~91,~265~(2010)}},
\texttt{\arxivref{0902.4458}{arxiv:0902.4458}}.

\bibitem{Gromov:2013pga}
N.~Gromov, V.~Kazakov, S.~Leurent and D.~Volin,
\textit{``{Quantum Spectral Curve for Planar AdS$_5$ /CFT$_4$ }''},
\textsf{\doiref{10.1103/PhysRevLett.112.011602}{Phys.Rev.Lett.~112,~011602~(2014)}},
\texttt{\arxivref{1305.1939}{arxiv:1305.1939}}.

\bibitem{Horigane:2009qb}
T.~Horigane and Y.~Kazama,
\textit{``{Exact Quantization of a Superparticle in $\mathit{AdS}_5 \times
  S^5$}''},
\textsf{\doiref{10.1103/PhysRevD.81.045004}{Phys.Rev.~D81,~045004~(2010)}},
\texttt{\arxivref{0912.1166}{arxiv:0912.1166}}.

\bibitem{Metsaev:1999gz}
R.~R.~Metsaev,
\textit{``{Light cone gauge formulation of IIB supergravity in
  ${\mathit{AdS}}_5 \times {S}^5$ background and AdS/CFT correspondence}''},
\textsf{\doiref{10.1016/S0370-2693(99)01063-1}{Phys.~Lett.~B468,~65~(1999)}},
\texttt{\arxivref{hep-th/9908114}{hep-th/9908114}}.

\bibitem{Metsaev:2000yu}
R.~R.~Metsaev, C.~B.~Thorn and A.~A.~Tseytlin,
\textit{``{Light-cone superstring in AdS space-time}''},
\textsf{\doiref{10.1016/S0550-3213(00)00712-4}{Nucl.~Phys.~B596,~151~(2001)}},
\texttt{\arxivref{hep-th/0009171}{hep-th/0009171}}.

\bibitem{Siegel:2010gm}
W.~Siegel,
\textit{``{Spacecone quantization of AdS superparticle}''},
\texttt{\arxivref{1005.5049}{arxiv:1005.5049}}.

\bibitem{Arvanitakis:2016vnp}
A.~S.~Arvanitakis, A.~E.~Barns-Graham and P.~K.~Townsend,
\textit{``{Anti-de Sitter particles and manifest (super)isometries}''},
\texttt{\arxivref{1608.04380}{arxiv:1608.04380}}.

\bibitem{Drukker:2000ep}
N.~Drukker, D.~J.~Gross and A.~A.~Tseytlin,
\textit{``{Green-Schwarz string in AdS(5) x S**5: Semiclassical partition
  function}''},
\textsf{\doiref{10.1088/1126-6708/2000/04/021}{JHEP~0004,~021~(2000)}},
\texttt{\arxivref{hep-th/0001204}{hep-th/0001204}}.

\bibitem{Frolov:2002av}
S.~Frolov and A.~A.~Tseytlin,
\textit{``Semiclassical quantization of rotating superstring in {$AdS_5 \times
  S^5$}''},
\textsf{\doiref{10.1088/1126-6708/2002/06/007}{JHEP~0206,~007~(2002)}},
\texttt{\arxivref{hep-th/0204226}{hep-th/0204226}}.

\bibitem{Berenstein:2002jq}
D.~Berenstein, J.~M.~Maldacena and H.~Nastase,
\textit{``Strings in flat space and pp waves from {$\mathcal{N}=\mathord{}$4}
  {Super} {Yang--Mills}''},
\textsf{\doiref{10.1088/1126-6708/2002/04/013}{JHEP~0204,~013~(2002)}},
\texttt{\arxivref{hep-th/0202021}{hep-th/0202021}}.

\bibitem{Gubser:2002tv}
S.~Gubser, I.~Klebanov and A.~M.~Polyakov,
\textit{``{A Semiclassical limit of the gauge / string correspondence}''},
\textsf{\doiref{10.1016/S0550-3213(02)00373-5}{Nucl.Phys.~B636,~99~(2002)}},
\texttt{\arxivref{hep-th/0204051}{hep-th/0204051}}.

\bibitem{Frolov:2003qc}
S.~Frolov and A.~A.~Tseytlin,
\textit{``Multi-spin string solutions in {$AdS_5\times S^5$}''},
\textsf{\doiref{10.1016/S0550-3213(03)00580-7}{Nucl.~Phys.~B668,~77~(2003)}},
\texttt{\arxivref{hep-th/0304255}{hep-th/0304255}}.

\bibitem{Arutyunov:2003uj}
G.~Arutyunov, S.~Frolov, J.~Russo and A.~A.~Tseytlin,
\textit{``Spinning strings in $AdS_5\times S^5$ and integrable systems''},
\textsf{Nucl.~Phys.~B671,~3~(2003)},
\texttt{\arxivref{hep-th/0307191}{hep-th/0307191}}.

\bibitem{Passerini:2010xc}
F.~Passerini, J.~Plefka, G.~W.~Semenoff and D.~Young,
\textit{``{On the Spectrum of the $AdS_5 \times S^5$ String at large
  lambda}''},
\textsf{\doiref{10.1007/JHEP03(2011)046}{JHEP~1103,~046~(2011)}},
\texttt{\arxivref{1012.4471}{arxiv:1012.4471}}.

\bibitem{Frolov:2013lva}
S.~Frolov, M.~Heinze, G.~Jorjadze and J.~Plefka,
\textit{``{Static gauge and energy spectrum of single-mode strings in AdS$_{5}
  \times$ S$^{5}$}''},
\textsf{\doiref{10.1088/1751-8113/47/8/085401}{J.Phys.~A47,~085401~(2014)}},
\texttt{\arxivref{1310.5052}{arxiv:1310.5052}}.

\bibitem{Jorjadze:2012iy}
G.~Jorjadze, J.~Plefka and J.~Pollok,
\textit{``{Bosonic String Quantization in Static Gauge}''},
\textsf{\doiref{10.1088/1751-8113/45/48/485401}{J.Phys.~A45,~485401~(2012)}},
\texttt{\arxivref{1207.4368}{arxiv:1207.4368}}.

\bibitem{Heinze:2015xxa}
M.~Heinze,
\textit{``{Spectrum and Quantum Symmetries of the ${\rm AdS}_5 \times {\rm
  S}^5$ Superstring}''},
\texttt{\arxivref{1507.03005}{arxiv:1507.03005}}.

\bibitem{Heinze:2014cga}
M.~Heinze, G.~Jorjadze and L.~Megrelidze,
\textit{``{Isometry Group Orbit Quantization of Spinning Strings in AdS$_{3}
  \times$ S$^3$}''},
\textsf{\doiref{10.1088/1751-8113/48/12/125401}{J.Phys.~A48,~125401~(2015)}},
\texttt{\arxivref{1410.3428}{arxiv:1410.3428}}.

\bibitem{Dzhordzhadze:1994np}
G.~Jorjadze, L.~O'Raifeartaigh and I.~Tsutsui,
\textit{``{Quantization of a relativistic particle on the SL(2,R) manifold
  based on Hamiltonian reduction}''},
\textsf{\doiref{10.1016/0370-2693(94)90549-5}{Phys.Lett.~B336,~388~(1994)}},
\texttt{\arxivref{hep-th/9407059}{hep-th/9407059}}.

\bibitem{Heinze:2015oha}
M.~Heinze, B.~Hoare, G.~Jorjadze and L.~Megrelidze,
\textit{``{Orbit method quantization of the AdS$_2$ superparticle}''},
\textsf{\doiref{10.1088/1751-8113/48/31/315403}{J.~Phys.~A48,~315403~(2015)}},
\texttt{\arxivref{1504.04175}{arxiv:1504.04175}}.

\bibitem{Heinze:2016fin}
M.~Heinze and G.~Jorjadze,
\textit{``{Quantization of the ${\rm AdS}_3$ superparticle on ${\rm
  OSP}(1|2)^2/{\rm SL}(2,\mathbb{R})$}''},
\textsf{\doiref{10.1016/j.nuclphysb.2016.11.018}{Nucl.~Phys.~B915,~44~(2017)}},
\texttt{\arxivref{1610.03519}{arxiv:1610.03519}}.

\bibitem{Zarembo:2010sg}
K.~Zarembo,
\textit{``{Strings on Semisymmetric Superspaces}''},
\textsf{\doiref{10.1007/JHEP05(2010)002}{JHEP~1005,~002~(2010)}},
\texttt{\arxivref{1003.0465}{arxiv:1003.0465}}.

\bibitem{Klose:2010ki}
T.~Klose,
\textit{``{Review of AdS/CFT Integrability, Chapter IV.3: N=6 Chern-Simons and
  Strings on AdS4xCP3}''},
\textsf{\doiref{10.1007/s11005-011-0520-y}{Lett.~Math.~Phys.~99,~401~(2012)}},
\texttt{\arxivref{1012.3999}{arxiv:1012.3999}}.

\bibitem{Sorokin:2011rr}
D.~Sorokin, A.~Tseytlin, L.~Wulff and K.~Zarembo,
\textit{``{Superstrings in AdS(2)xS(2)xT(6)}''},
\textsf{\doiref{10.1088/1751-8113/44/27/275401}{J.Phys.~A44,~275401~(2011)}},
\texttt{\arxivref{1104.1793}{arxiv:1104.1793}}.

\bibitem{Cagnazzo:2011at}
A.~Cagnazzo, D.~Sorokin and L.~Wulff,
\textit{``{More on integrable structures of superstrings in AdS(4) x CP(3) and
  AdS(2) x S(2) x T(6) superbackgrounds}''},
\textsf{\doiref{10.1007/JHEP01(2012)004}{JHEP~1201,~004~(2012)}},
\texttt{\arxivref{1111.4197}{arxiv:1111.4197}}.

\bibitem{David:2008yk}
J.~R.~David and B.~Sahoo,
\textit{``{Giant magnons in the D1-D5 system}''},
\textsf{\doiref{10.1088/1126-6708/2008/07/033}{JHEP~0807,~033~(2008)}},
\texttt{\arxivref{0804.3267}{arxiv:0804.3267}}.

\bibitem{Babichenko:2009dk}
A.~Babichenko, B.~Stefanski,~Jr. and K.~Zarembo,
\textit{``{Integrability and the $\mathit{AdS}_3/\mathit{CFT}_2$
  correspondence}''},
\textsf{\doiref{10.1007/JHEP03(2010)058}{JHEP~1003,~058~(2010)}},
\texttt{\arxivref{0912.1723}{arxiv:0912.1723}}.

\bibitem{Sfondrini:2014via}
A.~Sfondrini,
\textit{``{Towards integrability for ${\rm Ad}{{{\rm S}}_{{\bf 3}}}/{\rm
  CF}{{{\rm T}}_{{\bf 2}}}$}''},
\textsf{\doiref{10.1088/1751-8113/48/2/023001}{J.~Phys.~A48,~023001~(2015)}},
\texttt{\arxivref{1406.2971}{arxiv:1406.2971}}.

\bibitem{Galajinsky:2010zy}
A.~Galajinsky,
\textit{``{Particle dynamics near extreme Kerr throat and supersymmetry}''},
\textsf{\doiref{10.1007/JHEP11(2010)126}{JHEP~1011,~126~(2010)}},
\texttt{\arxivref{1009.2341}{arxiv:1009.2341}}.

\bibitem{Galajinsky:2011xp}
A.~Galajinsky and K.~Orekhov,
\textit{``{N=2 superparticle near horizon of extreme Kerr-Newman-AdS-dS black
  hole}''},
\textsf{\doiref{10.1016/j.nuclphysb.2011.04.015}{Nucl.Phys.~B850,~339~(2011)}},
\texttt{\arxivref{1103.1047}{arxiv:1103.1047}}.

\bibitem{Bellucci:2011hk}
S.~Bellucci and S.~Krivonos,
\textit{``{N=2 supersymmetric particle near extreme Kerr throat}''},
\textsf{\doiref{10.1007/JHEP10(2011)014}{JHEP~1110,~014~(2011)}},
\texttt{\arxivref{1106.4453}{arxiv:1106.4453}}.

\bibitem{Krivonos:2010zy}
S.~Krivonos and O.~Lechtenfeld,
\textit{``{Many-particle mechanics with $D(2, 1;\alpha)$ superconformal
  symmetry}''},
\textsf{\doiref{10.1007/JHEP02(2011)042}{JHEP~1102,~042~(2011)}},
\texttt{\arxivref{1012.4639}{arxiv:1012.4639}}.

\bibitem{Kozyrev:2013vla}
N.~Kozyrev, S.~Krivonos, O.~Lechtenfeld and A.~Nersessian,
\textit{``{Higher-derivative N=4 superparticle in three-dimensional
  spacetime}''},
\textsf{\doiref{10.1103/PhysRevD.89.045013}{Phys.~Rev.~D89,~045013~(2014)}},
\texttt{\arxivref{1311.4540}{arxiv:1311.4540}}.

\bibitem{Kozyrev:2016mlo}
N.~Kozyrev, S.~Krivonos and O.~Lechtenfeld,
\textit{``{Higher-derivative superparticle in AdS$_3$ space}''},
\textsf{\doiref{10.1103/PhysRevD.93.065024}{Phys.~Rev.~D93,~065024~(2016)}},
\texttt{\arxivref{1601.01906}{arxiv:1601.01906}}.

\bibitem{Galajinsky:2016wuc}
A.~Galajinsky and O.~Lechtenfeld,
\textit{``{Superconformal SU$(1, 1|n)$ mechanics}''},
\textsf{\doiref{10.1007/JHEP09(2016)114}{JHEP~1609,~114~(2016)}},
\texttt{\arxivref{1606.05230}{arxiv:1606.05230}}.

\bibitem{Fronsdal:1974ew}
C.~Fronsdal,
\textit{``{Elementary particles in a curved space. ii}''},
\textsf{\doiref{10.1103/PhysRevD.10.589}{Phys.~Rev.~D10,~589~(1974)}}.

\bibitem{Breitenlohner:1982jf}
P.~Breitenlohner and D.~Z.~Freedman,
\textit{``{Stability in Gauged Extended Supergravity}''},
\textsf{\doiref{10.1016/0003-4916(82)90116-6}{Annals~Phys.~144,~249~(1982)}}.

\bibitem{Aharony:1999ti}
O.~Aharony, S.~S.~Gubser, J.~M.~Maldacena, H.~Ooguri and Y.~Oz,
\textit{``Large N field theories, string theory and gravity''},
\textsf{Phys.~Rept.~323,~183~(2000)},
\texttt{\arxivref{hep-th/9905111}{hep-th/9905111}}.

\bibitem{Dorn:2005jt}
H.~Dorn and G.~Jorjadze,
\textit{``{On particle dynamics in AdS(N+1) space-time}''},
\textsf{\doiref{10.1002/prop.200510208}{Fortsch.Phys.~53,~486~(2005)}},
\texttt{\arxivref{hep-th/0502081}{hep-th/0502081}}.

\bibitem{Dorn:2005ja}
H.~Dorn and G.~Jorjadze,
\textit{``{Oscillator quantization of the massive scalar particle dynamics on
  AdS spacetime}''},
\textsf{\doiref{10.1016/j.physletb.2005.08.059}{Phys.Lett.~B625,~117~(2005)}},
\texttt{\arxivref{hep-th/0507031}{hep-th/0507031}}.

\bibitem{Dorn:2010wt}
H.~Dorn, G.~Jorjadze, C.~Kalousios and J.~Plefka,
\textit{``{Coordinate representation of particle dynamics in AdS and in generic
  static spacetimes}''},
\textsf{\doiref{10.1088/1751-8113/44/9/095402}{J.Phys.~A44,~095402~(2011)}},
\texttt{\arxivref{1011.3416}{arxiv:1011.3416}}.

\bibitem{Jorjadze:2012jk}
G.~Jorjadze, C.~Kalousios and Z.~Kepuladze,
\textit{``{Quantization of AdS x S particle in static gauge}''},
\textsf{\doiref{10.1088/0264-9381/30/2/025015}{Class.Quant.Grav.~30,~025015~(2013)}},
\texttt{\arxivref{1208.3833}{arxiv:1208.3833}}.

\bibitem{Heinze:footnote1}
{More precisely, $J_{AB}$ are even covariant under the compact subgroup
  $\SO(2)\times\SO(N) \subset \SO(2,N)$}.

\bibitem{Heinze:footnote2}
{We assume $N\geq 3$, such that he operator $J_N$ is positive and its inverse
  is well defined. Note that the cases $N=1$ and $N=2$ reduce to the
  Holstein-Primakoff representation, see Appendix \ref{app:bosAdS23}}.

\end{thebibliography}

\end{document}